\documentclass[twocolumn,nofootinbib,superscriptaddress,
               showpacs,preprintnumbers,
               amsmath,amssymb]
              {revtex4}
\usepackage{graphicx}

\newcommand{\e}{\mathrm{e}}
\newcommand{\Imag}{\mathop{\mathrm{Im}}}
\newcommand{\Real}{\mathop{\mathrm{Re}}}
\newcommand{\bra}[1]{\langle #1 |}
\newcommand{\ket}[1]{| #1 \rangle}

\begin{document}

\preprint{INR/TH-2003-1}

\title{Transmission through a potential barrier in quantum mechanics of
  multiple degrees of freedom: complex way to the top.}

\author{F.~Bezrukov}
\email{fedor@ms2.inr.ac.ru}
\affiliation{
  Institute for Nuclear Research of the Russian Academy of Sciences,\\
  60th October Anniversary prospect 7a, Moscow 117312, Russia}
\author{D.~Levkov}
\email{levkov@ms2.inr.ac.ru}
\affiliation{
  Institute for Nuclear Research of the Russian Academy of Sciences,\\
  60th October Anniversary prospect 7a, Moscow 117312, Russia}
\affiliation{
  Moscow State University, Department of Physics,\\
  Vorobjevy Gory, Moscow, 119899, Russian Federation}
\date{January 7, 2003}

\begin{abstract}
  A semiclassical method for the calculation of tunneling exponent in
  systems with many degrees of freedom is developed.  We find that
  corresponding classical solution as function of energy form several
  branches joint by bifurcation points.  A regularization technique is
  proposed, which enables one to choose physically relevant branches
  of solutions everywhere in the classically forbidden region and also
  in the allowed region.  At relatively high energy the physical
  branch describes tunneling via creation of a classical state, close
  to the top of the barrier.  The method is checked against exact
  solutions of the Schr\"odinger equation in a quantum mechanical
  system of two degrees of freedom.
\end{abstract}

\pacs{03.65.Sq, 
      03.65.-w, 
      03.65.Xp
     }

\maketitle

\onecolumngrid
\tableofcontents
\twocolumngrid
\mbox{}
\cleardoublepage

\section{\label{sec:intro}Introduction}

Semiclassical methods provide a very useful tool for the study of
nonperturbative processes.  Tunneling phenomena represent one of the
most notable cases where semiclassical techniques can be used to
obtain otherwise unattainable information on transition probabilities.
There are many cases where the application of semiclassical methods
reduces to having to solve the classical equations of motion in the
Euclidean time domain, namely with $t$ analytically continued to
$i\tau$.  A famous example is WKB approximation to tunneling in
quantum mechanics with one degree of freedom.  In systems with
multiple degrees of freedom there are other, more complex situations,
however, where the evolution along the imaginary time axis cannot be
treated separately from the real time evolution which precedes and
follows it.  This happens, for example, in the study of tunneling in
high-energy collisions in field theory
\cite{Kuznetsov:1997az,Bezrukov:2001dg}, where one must consider a
system with definite particle number in the initial state, or in the
study of chemical reactions where the initial system is in a definite
quantum state \cite{Miller}.  Then, in the search for complex-time
solutions of the classical equations of motion that dominate
tunneling, one encounters a novel phenomenon which, if not properly
handled, may make it impossible to obtain the appropriate solutions.
We ran into this kind of problems in the study of an interesting class
of field theoretical processes, where tunneling occurs between
different topological sectors and is accompanied by baryon number
violation
\cite{'tHooft:1976fv,Ringwald:1990ee,Espinosa:1990qn,Bezrukov:2001dg}.
However, the specifics of the model are immaterial for the features of
tunneling which we plan to describe in this paper.  Indeed, the fact
that, with a field theoretical system, one is dealing with an infinite
number of degrees of freedom is also irrelevant for the appearance of
the phenomenon that we alluded to above: this phenomenon emerges as
soon as one deals with more than one degree of freedom and in quite a
range of applications of semiclassical techniques.  For this reason we
decided to illustrate it in this paper in the context of a simple
quantum mechanical system with two degrees of freedom. The problems
one encounters in the study of the field theoretical system and their
solution will be described in a separate paper \cite{BLRRT:prepare}.
Our hope is that the discussion presented here may help to shed some
light on how semiclassical methods can be effectively applied in
situations which a priory may seem intractable.

\begin{figure}
  \centerline{\includegraphics[width=0.8\columnwidth]{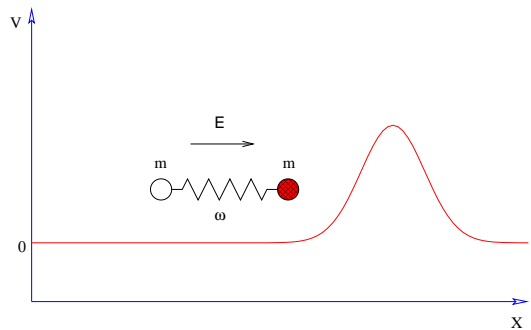}}
  \caption{Oscillator hitting a potential barrier, with which only the
 ``dark'' particle interacts.}
  \label{fig0}
\end{figure}

\begin{figure}
\centerline{\includegraphics[width=\columnwidth]{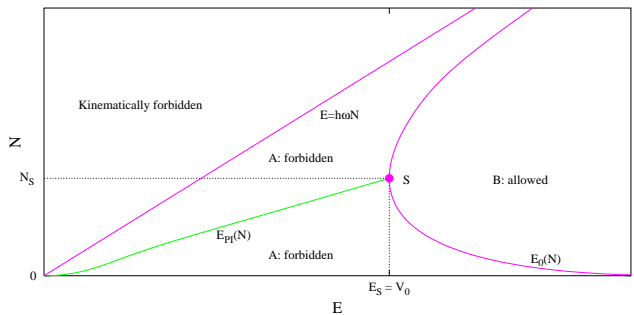}}
  \caption{Allowed and forbidden regions in the model with two
    degrees of freedom.}
\label{fig1}
\end{figure}

Our considerations are of quite general validity, but, for purposes of
clarity, it is useful to present them in the context of a definite
model.  We will take this to be the model of
Ref.~\cite{Bonini:1999kj,Bonini:1999cn}, namely a system formed by
two particles of identical mass $m$, moving in one dimension and bound
by a harmonic oscillator potential of frequency $\omega$ (Fig.~\ref{fig0}).
One of the
particles interacts with a repulsive potential barrier.  This potential barrier
 is
assumed to be high and wide, while the spacing between the oscillator
levels is much smaller than the barrier height $V_0$.  Thus, the
Hamiltonian of the model is
\begin{equation}\label{Hold}
  H = \frac{p_1^2}{2m}+\frac{p_2^2}{2m}+\frac{m\omega^2}{4}(x_1-x_2)^2
      +V_0\e^{- x_1^2/2\sigma^2} \;,
\end{equation}
where the conditions on the oscillator frequency and potential barrier are
\begin{eqnarray}
  &&\hbar\omega\ll V_0, \label{conditions}\\
  &&\sigma\gg\hbar/\sqrt{mV_0}\;.\nonumber
\end{eqnarray}
The problem is to calculate the transition probability through the
barrier for a given incoming bound state; since the variables do not
separate, this is certainly a non-trivial
problem. Quantum-mechanically the incoming state is fully
characterized by its total energy $E$ and oscillator excitation number
$N$.  For $E<V_0$ the transition through the barrier can happen only
through tunneling.  However, the transition may be classically
forbidden, and occur through tunneling, also if $E>V_0$.  A detailed
study of the model (Ref.~\cite{Bonini:1999kj,Bonini:1999cn}) shows
that one can distinguish two regions in the $E-N$ plane, as
illustrated in Fig.~\ref{fig1}: region A where transition across the
barrier is classically forbidden, and region B where it is classically
allowed.  (Since $\hbar\omega (N-\frac{1}{2})$ cannot exceed $E$, the
further region with $N > E/\hbar\omega-\frac{1}{2}\simeq
E/\hbar\omega$ is trivially excluded.)  The vertex S of B marks the
lowest energy at which one can have a classically allowed transition,
characterized by $E=E_S=V_0$ and some definite initial excitation
number $N=N_S$.  Strictly speaking, in order to have a classically
allowed transition the energy must be (infinitesimally) larger than
$E_S$.  For $E=E_S$ there exists a static, unstable solution to the
classical equations of motion, where the system just sits on top of
the barrier: $x_1=x_2=0$.  In the study of field theoretical tunneling
processes this solution has been called \emph{the
sphaleron}~\cite{Klinkhamer:1984di} and, for sake of terminology, we
will use the term also in this article, but no special meaning should
be attributed here to it: for us it will be just a name.  The line
$E_{\mathrm{PI}}$ in Fig.~\ref{fig1} marks a special class of
tunneling solutions to the equations of motion, about which more will
be said later.

In the region A one would like to calculate the exponent $F(E,N)$ in
the transition probability $\mathcal{T}\approx\exp(- F/\hbar)$.  The
boundary between regions A and B is given by the line $E_0(N)$, where
$F(E,N)=0$.  $F$ can be calculated by the method introduced in
Ref.~\cite{Rubakov:1992ec}, which will be recapitulated later. (See
also the earlier work by Miller~\cite{Miller} for an elegant approach
to the calculation of the transition probability, based on canonical
transformations.)  The method of Ref.~\cite{Rubakov:1992ec} leads to
having to solve the classical equations of motion, after analytic
continuation to complex time and complex phase space, subject to
suitable boundary conditions.  In the tunneling solutions, the system
evolves along the real time direction first, then undergoes the
under-barrier evolution along the imaginary time direction.  For the
solutions situated on the line $E_{\mathrm{PI}}$ in Fig.~\ref{fig1},
the coordinates of the particles are real throughout the evolution,
but this is a very special case.  For general tunneling solutions the
coordinates and the corresponding momenta, have to be continued to
complex values.  This is a defining feature of processes with more
than one degree of freedom: in systems where a single particle moves
along a line and tunnels under a potential barrier, all tunneling
solutions would be real and behave like the processes on the line
$E_{\mathrm{PI}}$. In systems with multiple degrees of freedom
tunneling solutions, i.e.\ evolutions which obey the classical
(analytically continued) equations of motion and obey the appropriate
boundary conditions, can be rather readily found for $E<E_S$.  As the
energy approaches the barrier height, however, something special
happens: in the tunneling solutions the system lingers for
progressively longer times on top of the barrier and the boundary
conditions become truly asymptotic in time.  Rigorously, finite time
evolutions that satisfy the boundary conditions cease to exist.  Here
lies the subtlety in the application of semiclassical methods to such
processes and the root of the difficulty in finding solutions that can
be used throughout the domain of classically forbidden transitions
(i.e.~the whole of region A).  As we plan to show in this paper, the
difficulty can be overcome by employing a suitable regularization, and
one can find physically relevant solutions which map the entire region
of classically forbidden processes and also merge continuously with
classically allowed evolutions as one approaches the boundary
$E_0(N)$.  The new solutions we find all describe tunneling onto the
top of the potential barrier, or onto long-living excitations above
the top. Notice that the problem is special to transitions which have
an energy larger than the barrier height (i.e.\ larger than the
sphaleron energy) and thus would never be encountered in systems with
a single degree of freedom, the systems normally used in textbooks to
illustrate the use of semiclassical techniques, where all processes
with energy larger than the barrier height are classically allowed.
Still, even in simple systems with a single degree of freedom, one can
see a sign of the subtleties that one can, and does encounter in more
complex systems.  For this reason, and also in order to put our whole
formulation on a clear footing, we will start by re-examining the
semiclassical description of tunneling in the case of a single
particle moving on a line.  This will be done in Sec.~\ref{sec:1D}.

In Sec.~\ref{sec:2D} we consider the quantum mechanical model with two
degrees of freedom. In particular, in Sec.~\ref{sec:Ttheta} we
introduce the semiclassical technique for the calculation of the
tunneling exponent. Then we examine the classical over-barrier
solutions and find the boundary of the classically allowed region
$E_0(N)$ in Sec.~\ref{sec:over-barrier}.  In
Sec.~\ref{sec:bifurcation} we present a straightforward application of
the semiclassical technique, outlined in Sec.~\ref{sec:Ttheta}, and
find that it ceases to produce the relevant classical solutions in a
part of the tunneling region~A, roughly speaking, at $E>E_\mathrm{S}$.
In Sec.~\ref{sec:4} we introduce our regularization technique and show
that it indeed enables one to find the tunneling classical solutions
in the entire region~A (Sec.~\ref{sec:reg_forb}).  We check our method
against the numerical solution of the full Schr\"odinger equation in
Sec.~\ref{sec:4.2}.  In Sec.~\ref{sec:4.3} we show how our
regularization technique may be used to join smoothly the classically
allowed and classically forbidden regions.  Our conclusions are
presented in Sec.~\ref{sec:conclusions}.

\section{\label{sec:1D}Quantum mechanics of one degree of freedom}

\subsection{\label{sec:1D.1}Classical boundary value problem}

Let us review the derivation of the semiclassical expression for the
tunneling exponent in quantum mechanics of one degree of freedom.  The
result is well known in this case (WKB formula) and the derivation
here may seem artificially complicated, but it can be generalized to
the case of multiple degrees of freedom, while many of its general
features can still be captured in the relatively simple case of one
degree of freedom.

Let us calculate the probability of tunneling through a potential
barrier, located in a region near  $X=0$, from the asymptotic region
$X\to-\infty$.  Let us use a system of
units with $\hbar=1$ and set the mass of a particle equal to 1
(both can be reintroduced in our formulas on dimensional grounds).
The Hamiltonian we consider has the general form
\begin{equation}\label{L1dim}
  \hat{H}=\left(\frac{\hat{P}^2}{2}+V(\hat{X})\right)\;,
\end{equation}
where the potential $V(X)$ has the form
\begin{equation*}
V(X) = \frac{1}{\lambda} U(\sqrt{\lambda} X)
\end{equation*}
and rapidly decays as $X\to \pm\infty$.
Then the semiclassical limit of this model is obtained when $\lambda\to0$
while the energy is kept parametrically large: $E=\tilde{E}/\lambda$. 

The probability of tunneling from an incoming ($X\to - \infty$) state 
with energy $E$ is
\begin{equation}
  \mathcal{T}(E) = \lim_{t_f - t_i \to \infty}
    \int\limits_0^{+\infty} dX_f
    \left|\bra{X_f}\e^{-i\hat{H}(t_f - t_i)}\ket{E}\right|^2
  \;,\label{*}
\end{equation}
where it is implicit that the initial state $|E\rangle$ is a wave packet with
very small spread in momentum $P$, centered at large negative $X$. 
The transition amplitude 
\begin{equation*}
  \mathcal{A}_{fi} = \bra{X_f} \e^{-i \hat{H}(t_f - t_i)} \ket{X_i}
\end{equation*}
and its complex conjugate $\mathcal{A}_{i'f}^*$ have the standard path
integral representation
\begin{eqnarray}\label{AA}
  \mathcal{A}_{fi} &=& \int [dX]
  \Bigg|_{\stackrel{
    \scriptscriptstyle X(t_i) = X_i}{
    \scriptscriptstyle X(t_f) = X_f}}
  \e^{iS[X]}
  \;,\\
  \mathcal{A}_{i'f}^* &=& \int [dX']\Bigg|_{\stackrel{
    \scriptscriptstyle{X}'(t_i) =  X'_i}{
    \scriptscriptstyle{X}'(t_f) =  X_f}}
  \e^{-iS[X']}
  \;, \nonumber
\end{eqnarray}
where $S[X]=\int_{t_i}^{t_f}L(X,\dot{X})dt$ is the action of the model. 
Inserting factors $\int dX_i \;|X_i\rangle\langle X_i| = 1$ one 
re-expresses \eqref{*} in terms of these transition amplitudes and the initial 
state matrix elements
\begin{equation*}
  {\cal B}_{ii'} = \langle X_i| E\rangle
  \langle E| X_i'\rangle
\end{equation*}
in the following way,
\begin{equation}
  \mathcal{T}(E) = \lim\limits_{t_f-t_i \to \infty}
    \int\limits_{0}^{+\infty} dX_f \int\limits_{-\infty}^0 d{X}_id{X}_i'
    \, \mathcal{A}_{fi} \mathcal{A}_{i'f}^*\mathcal{B}_{ii'} .\label{TTT*}
\end{equation}
To obtain an integral representation of the initial-state matrix elements,
let us rewrite ${\cal B}_{ii'}$ as follows,
\begin{equation}\label{Bii}
  {\cal B}_{ii'} = 
  \langle X_i| \hat{P}_E|X_i'\rangle\;,
\end{equation}
where $\hat{P}_E$ denotes the projector onto a state with total energy
$E$.  In the momentum basis this projector has an integral
representation which follows from the integral representation of the 
$\delta$--function:
\begin{equation}\label{PEexpr}
  \bra{Q} \hat{P}_E \ket{P} = \frac{\sqrt{2 E}}{2 \pi}
  \int\! d\xi\, \e^{ - i (E - \frac{P^2}{2})\xi}\,\delta(Q - P)\;,
\end{equation}
where $\ket{Q}$, $\ket{P}$ are the momentum eigenstates.
It is convenient to introduce the notation
\begin{equation*}
  T = - i \xi\;.
\end{equation*}
Combining formulas \eqref{AA}, \eqref{TTT*}, \eqref{Bii},
\eqref{PEexpr}, and evaluating the Gaussian integrals over $P$ and
$Q$, one obtains the path integral representation for the tunneling
probability.  After rescaling $X\to X/\sqrt{\lambda},\; X'\to
X'/\sqrt{\lambda}$ it takes the following form
\begin{multline}\label{**}
  \mathcal{T}(E) = \lim\limits_{t_f-t_i \to
    \infty}\Bigg\{ \int\limits_{-i\infty}^{+i\infty}  dT
    \int [dX\,dX'] \\
    \exp\left[- \frac{1}{\lambda}F[X,X';T]\right]\Bigg\},
\end{multline}
where the integration over $X_i$, $X'_i$ and $X_f$ is implicit, and
\begin{multline}\label{F1dim}
  F[X,X';T] = - i S[X] + i S[X']\\
    - \tilde{E} T + \frac{(X_i - X_i')^2}{2 T}
  \;,
\end{multline}
with $\tilde{E}=\lambda E$ being the rescaled energy (independent of $\lambda$
in the semiclassical limit $\lambda\to0$).  In this paper
we will omit the tilde over the rescaled parameters, where this does not
cause confusion.  

\begin{figure}
  \centerline{\includegraphics[width=\columnwidth]{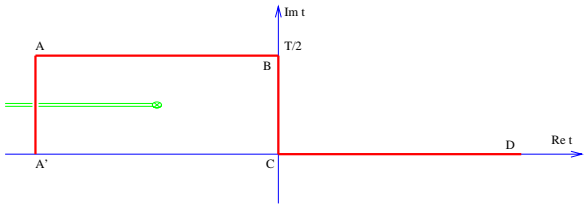}}
  \caption{Contour in the complex time plane.}
\label{fig2}
\end{figure}

At small $\lambda$  one applies the saddle point approximation to 
the  integral~\eqref{**}.  We will not consider the pre-exponential factor
in the probability, so we will be interested in the saddle point value of $F$. 
The saddle point equations for $X(t)$ and $X'(t)$ are the classical equations 
of motion,
\begin{equation}\label{+}
  \frac{\delta S(X)}{\delta X} = 0 
  \;,\quad
  \frac{\delta S(X')}{\delta X'}= 0
  \;,
\end{equation} 
while the variation with respect to the boundary values $X_f$, $X_i$, $X'_i$ 
and auxiliary variable $T$ leads to the boundary conditions
\begin{subequations}
\label{7}
\begin{eqnarray}
 && P_f = P'_f\label{7a}\;,\\
 && P_i = P'_i =- \frac{X_i-X'_i}{iT}\;,\label{1dbcin}\\
 && E = -\frac{(X_i-X'_i)^2}{2T^2}\label{1dbcout}\;,
\end{eqnarray}
where $P\equiv \dot{X}$. Note also, that one has~(see Eq.~\eqref{AA})
\begin{equation}
X(t_f) = X'(t_f) = X_f \label{7d}\;,
\end{equation}
\end{subequations}
and that Eq.~\eqref{1dbcout} may be written in the form $E = P_i^2/2$.
The latter equation requires that the energy of the solution is equal to $E$. 
One more observation is that
\begin{equation}
X' = X^*\label{++}\;,
\end{equation}
because $X'$ corresponds to the saddle point of the complex conjugate
amplitude.  Equations~\eqref{7a} and~\eqref{7d} simply mean that the
solutions $X(t)$ and $X'(t)$ coincide as $t\to+\infty$.  On the other
hand, Eq.~\eqref{1dbcin} states that the solutions $X(t)$ and $X'(t)$
are different at initial time $t\to -\infty$, if the saddle point
value of $T$ is not zero.  There is no contradiction: for energy $E$
larger than the barrier height, a real classical trajectory running
from $X\to-\infty$ to $X\to+\infty$ exists; for this solution $T=0$
and $X$ coincides with $X'$ at any moment of time.  On the other hand,
if $E$ is smaller than the barrier height, then there is no such
classical solution in real time, but the solution may exist on a
contour in the complex time plane, which wraps around some branch
point (as shown in Fig.~\ref{fig2}). The solution continued along this
contour is in general complex at large negative times (region A' in
Fig.~\ref{fig2}).  Then equations \eqref{1dbcin}, \eqref{1dbcout} and
\eqref{++} guarantee that the saddle point value of $T$ is real.

The boundary conditions may be significantly
simplified by shifting the initial time from the real axis.  
As $\Real t \to - \infty$, the potential is irrelevant,  
and any solution describes free motion with momentum $P_i = \sqrt{2 E}$,
\begin{eqnarray}
&&  X(t) = X_i+P(t-t_i)\;,\nonumber\\
&&  X'(t) = [X(t)]^*\;.\nonumber
\end{eqnarray}
Let us choose the initial time in such a way that $\Imag t = T/2$, 
i.e.~consider the boundary value problem~\eqref{+},~\eqref{7} on the contour
 ABCD
of Fig.~\ref{fig2}.
Then Eq.~\eqref{1dbcin} takes a simple form
\begin{eqnarray*}
  && \frac{X^*(t)-X(t)}{i T}=0 \,,\\
  && t = t'+\frac{iT}{2} \,,\quad t'=\text{real}\to-\infty\;,
\end{eqnarray*}
which simply means that the coordinate is real on the part AB of the
time contour. Equation~\eqref{++} implies that the solution is real in
the asymptotic region D, i.e.\ at $t = \text{real}\to +\infty$.
Hence, instead of Eq.~\eqref{7d} one has reality conditions in the
asymptotics A and D.  After obtaining the solution to the classical
boundary value problem, one inserts it to the right hand side of
Eq.~\eqref{F1dim} and obtains
\begin{equation}\label{Fanswer}
  F(E) = 2\Imag S-ET \;,
\end{equation}
where $S$ is the classical action on the contour ABCD.

\begin{figure}
  \begin{center}
    \includegraphics[width=0.9\columnwidth]{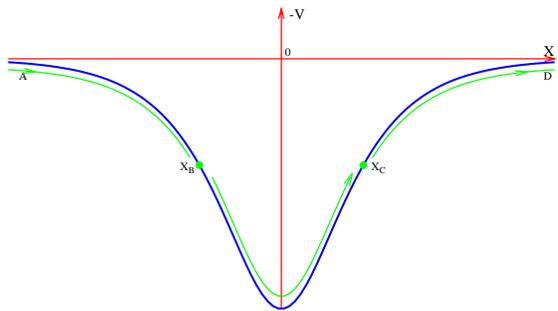}
  \end{center}
  \caption{Classical solution describing tunneling.}
  \label{fig3}
\end{figure}

In quantum mechanics of one degree of freedom, the contour ABCD may be
chosen in such a way that the points B and C are the turning points of
the solution.  Then the solution is real also at the part BC of the
contour.  Indeed, a real solution at the part BC of the contour
oscillates in the upside-down potential, $T/2$ equals to half-period
of oscillations, and the points B and C are the two different turning
points, $\dot{X} = 0$ (see Fig.~\ref{fig3}).
The continuation of this solution, according to the equation
of motion, from the point C to the positive real times corresponds to the 
real-time motion, from the point $X_C$ with zero initial velocity,
towards $X\to+\infty$; the coordinate $X(t)$ stays real on the part CD of the 
contour. Likewise, the continuation back in time from the point C leads to real
solution in the part AB of the contour. In this way the reality conditions 
are satisfied at A and D. 
The only contribution to $F$
comes from the Euclidean part of the contour, and 
one can check that the expression~\eqref{Fanswer} reduces to 
\begin{equation}\label{Fanswer1}
  F(E) = 2 \int\limits_{X_B}^{X_C}\! \sqrt{2(V(X)-E)}\, dX \;,
\end{equation}
which is the standard WKB result.

To summarize, in quantum mechanics of one degree of freedom, the semiclassical 
calculation of the tunneling exponent may be performed by solving the 
classical equation of motion on the contour ABCD in complex time plane,
with the conditions that the solution be real in the asymptotic past (region
 A),
and asymptotic future (region D). The relevant solutions tend to $X\to -\infty$
and $X\to + \infty$ in regions A and D, respectively. The auxiliary parameter 
$T$
is related to the energy of incoming state by the requirement that the
energy of the classical solution equals to $E$. The exponent for
the transition probability is given by Eq.~\eqref{Fanswer}, which, in fact, 
coincides with the WKB result~\eqref{Fanswer1}.

\subsection{\label{sec:1D.2}Joining the allowed and forbidden regions}

The solutions appropriate in the classically forbidden and classically
allowed regions apparently belong to quite different branches. It is
clear from Fig.~\ref{fig3} that as the energy approaches the height of
the barrier $V_0$ from below, the amplitude of the oscillations in the
upside-down potential decreases, while the period $T$ tends to a
finite value determined by the curvature of the potential at its
maximum. On the other hand, the solutions for $E > V_0$ (classically
allowed region) always run along the real time axis, so the parameter
$T$ is always zero. Hence, the relevant solutions do not merge at $E =
V_0$, and $T(E)$ has a discontinuity at $E = V_0$. We find it
instructive to develop a regularization technique that removes this
discontinuity and allows for smooth transition through the point $E =
V_0$. The reason is that in quantum mechanics of multiple degrees of
freedom, similar points exist not only at the boundary of the
classically allowed region, but also well inside the forbidden region
(but still at $E > E_S$, see Introduction and
Sec.~\ref{sec:over-barrier}).

To illustrate the situation, let us consider an exactly solvable model with 
\begin{equation*}
U(X) = \frac{1}{\cosh^2 X}\;.
\end{equation*}

The solution of the classical boundary value problem for energy smaller than 
the barrier height, $E<V_0\equiv 1$, is
\begin{equation}\label{1dtunsol}
  \sqrt{\frac{E}{1-E}}\;\sinh X = -\cosh\left(\sqrt{2E}(t-t_0)\right)
  \;.
\end{equation}
This solution has branch points at $t^\pm_*$, where the potential
becomes singular ($\cosh X=0$):
\begin{equation}\label{1dimsolunder}
  t^\pm_*-t_0 = \frac{1}{\sqrt{2E}}\left(
    \pm \log\frac{1+\sqrt{E}}{\sqrt{1-E}}
    +i\frac{\pi}{2}+i\pi n
  \right)\;.
\end{equation}
The requirement that the solution is real
on the part AB of the contour determines the imaginary part of $t_0$,
\begin{equation}\label{Imt0}
  \Imag t_0 = T/2\;.
\end{equation}
To fix the translational invariance along real time, one imposes a
constraint on $\Real t_0$.  The solutions that are real at imaginary 
(Euclidean) time are obtained for 
$$
\Real t_0 = 0\;.
$$
Later on, another choice will be convenient, for which the 
real parts of the position of the left singularities
($t^-_*$) are held  at a certain negative value $-c$.  This gives
\begin{equation}\label{Ret0}
  \Real t_0 = \frac{1}{\sqrt{2E}}\log\frac{1+\sqrt{E}}{\sqrt{|1-E|}}-c
  \;.
\end{equation}
We require $X$ to be real on the CD part of the contour (in general,
only asymptotically), and in this way determine the value of $T/2$.
In the classically forbidden case, $E<1$,  we are interested in  two branches of solutions.
The first branch has $T/2=0$, i.e.\ the contour ABCD reduces to real axis; the
classical solutions on this branch approach the barrier from the left and 
bounce back to $X\to - \infty$. This branch is not relevant for tunneling. 
The tunneling branch\footnote{There is in fact an infinite
  set of tunneling solutions with $T/2=(2n+1)\pi/\sqrt{2E}$ and
  reflecting solutions with $T/2=2n\pi/\sqrt{2E}$, but the imaginary
  part of their action is larger than that of the solutions we
  discuss, so their contribution to tunneling is exponentially
  small.} has
\begin{equation}\label{TE}
  \frac{T}{2}=\frac{\pi}{\sqrt{2E}} \;,
\end{equation}
Equations \eqref{Imt0}, \eqref{Ret0} and \eqref{TE}
determine the tunneling solution completely, while the transmission
exponent is obtained by plugging this solution into
Eq.~\eqref{Fanswer}.

It is straightforward to obtain the classical solutions at $E>1$, 
\begin{equation}\label{1dimsolover}
  \sqrt{\frac{E}{E-1}}\;\sinh X = \sinh\left(\sqrt{2E}(t-t_0)\right)
  \;.
\end{equation}
The role of the two branches is now reversed. The branch with $T/2 = 0$
describes the over--barrier transitions, i.e., it is the relevant one. 
The branch with $T/2 = \pi/\sqrt{2 E}$ still exists, but it has $X\to - \infty$
at both part A and part D of the contour, and thus is irrelevant.
We note in passing that all these solutions are real on the CD part of the 
contour
not only asymptotically, but at finite times also.

It is seen that the transition from the tunneling to the over-barrier
solutions is indeed not smooth: $T$ has a  discontinuity at the energy equal to
barrier height $E=V_0\equiv 1$. 
The energy $E = V_0\equiv 1$ is special also in the following respect. 
As the energy tends to $V_0$ from below, the relevant classical solutions
(with $T/2 = \pi/\sqrt{2 E}$) stay longer and longer near the top of the 
barrier. In the limit $E\to 1$ the time the solution spends near $X = 0$
is infinitely long. This is reflected also in Eq.~\eqref{1dimsolunder}: the 
distance between the branch points 
\begin{equation}\label{+'}
  t_*^+ - t_*^- = \sqrt{\frac{2}{E}} \log\frac{1 + \sqrt{E}}{\sqrt{|1-E|}}
\end{equation}
tends to infinity as $E\to 1$. The same situation occurs when the energy
approaches the height of the barrier from above: the classical over--barrier
solutions get ``stuck'' near $X = 0$. 
In a certain sense, all these solutions tend, as $E\to V_0\equiv 1$, to the
unstable static solution $X(t) = 0$---the sphaleron---describing a particle 
that sits on the top of the potential barrier. The latter property suggests a 
method to deal with this discontinuous behavior of the solutions with
respect to energy.
The idea is to regularize the classical boundary value problem in such a way
that the classical solution $X(t) = 0$ be unreachable. We implement this idea
by formally changing the potential
\begin{equation}\label{Uintnew1D}
  U(X) \to \e^{- i \epsilon} U(X)
  \;,
\end{equation}
which leads to a corresponding change of the classical equations of
motion. Here $\epsilon$ is a real regularization parameter, the smallest 
parameter in the model. At the end of the calculations one takes the limit 
$\epsilon\to 0$.

We do not change the boundary conditions in our classical problem,
i.e., we still require that $X(t)$ be real in the asymptotic future on
the real time axis and that $X(t')$ be real as $t'\to -\infty$ on the
part A of the contour ABCD. Then the conserved energy is real. The
sphaleron solution $X(t) = 0$ has now \emph{complex} energy (because
the potential is complex). Hence, the solutions to our classical
boundary value problem necessarily avoid the sphaleron, and one may
expect that the solutions behave smoothly in energy.

We will have to say more about our regularization in Sec.~\ref{sec:4}:
here we show that it indeed does the job of connecting the
over--barrier solutions smoothly to the tunneling ones.

The general solution to the regularized problem can be obtained from
\eqref{1dtunsol},
\begin{equation*}
  \sqrt{\frac{E}{\e^{-i \epsilon }-E}}\sinh X
  = - \cosh\left(\sqrt{2 E}(t-t_0)\right)\;,
\end{equation*}
where $t_0$ is the integration constant.  The value of $\Imag t_0$ is fixed
by the requirement that $\Imag X=0$ at positive time $t\to +\infty$,
\begin{equation*}
  \Imag t_0= \frac{T}{2} - \frac{1}{2\sqrt{2E}}
\mathrm{arg}[\mathrm{e}^{-i\epsilon} - E]\;.
\end{equation*}
The condition that the coordinate $X$ is real on the initial part AB
of the contour gives the relation between $T$ and $E$,
\begin{equation}\label{TEThetaE1D}
  \frac{T}{2} = \frac{1}{\sqrt{2E}}\left\{
    \pi+\arg\left(\mathrm{e}^{-i\epsilon}-E\right) \right\}\;.
\end{equation}
For $\epsilon=0$ and $E<1$, the result $T/2= \pi/\sqrt{2E}$ is
reproduced.

\begin{figure}
  \begin{center}
    \includegraphics[width=0.9\columnwidth]{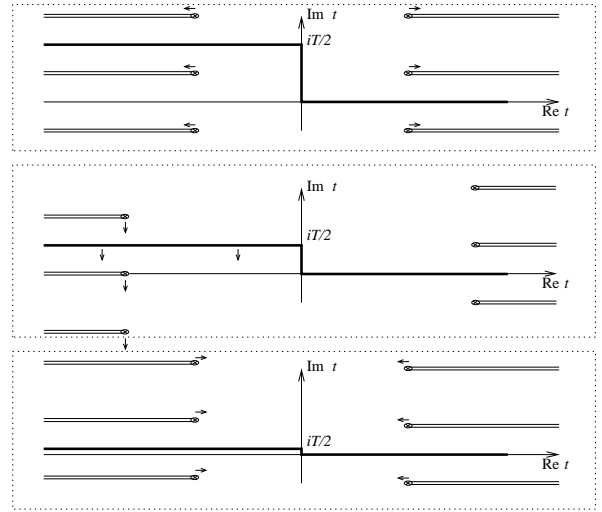}
  \end{center}
  \caption{The contour in complex time plane for $\epsilon  \ne 0$,
    $E < 1$ (upper figure), $E = 1$ (the middle figure)
    and $E > 1$ (lower figure). Arrows show the motion of the contour
and branch points as energy increases.}
\label{fig4}
\end{figure}

\begin{figure}
\begin{center}
  \includegraphics[width=\columnwidth]{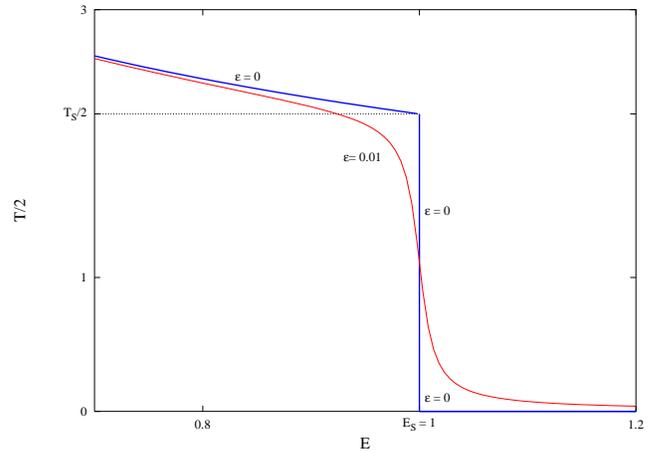}
\end{center}
  \caption{$T$ as function of $E$ for zero and 
nonzero $\epsilon $, thick and thin lines, respectively.}
\label{fig5}
\end{figure}

Let us analyze what happens in the regularized case in the vicinity
of the would-be special value of energy, $E = 1$. It is clear from 
Eq.~\eqref{TEThetaE1D} that T is now a 
smooth function of $E$. Away from $E = 1$,
Eq.~\eqref{TEThetaE1D} can be written as follows,
\begin{equation}\label{TEthetaE_small}
  \frac{T}{2} 
  = \left\{\begin{array}{ll}
      \displaystyle\frac{\pi}{\sqrt{2E}} & \text{forbidden region, }
1-E \gg \epsilon \\
      \displaystyle\frac{\epsilon }{\sqrt{2 E}(E-1)} & 
\text{allowed region, }  E-1 \gg \epsilon.
    \end{array}\right.\hspace{-0.7em}
\end{equation}
Deep enough in the forbidden region, when $1-E\gg \epsilon$, the
argument in equation~(\ref{TEThetaE1D}) is nearly zero and we return
to the original result~\eqref{TE}.  When $E$ crosses the region of
size of order $\epsilon$ around $E = 1$, the argument rapidly changes
from $O(\epsilon)$ to $-\pi$, so that $T/2$ changes from
$\pi/\sqrt{2}$ to nearly zero.  Thus, at $E > 1$ we arrive to a
solution which is very close to the classical over-barrier transition,
and the contour is also very close to the real axis.  This is shown in
Figs.~\ref{fig4},~\ref{fig5}. We conclude that at small but finite
$\epsilon$, the classically allowed and classically forbidden regions
merge smoothly.

At ${E<1}$, the limit $\epsilon \to 0$ is straightforward.  For
${E>1}$ a somewhat more careful analysis of the limit ${\epsilon
\to0}$ is needed.  From Eq.~\eqref{TEthetaE_small} one observes that
the limit ${\epsilon \to0}$ with constant finite $T<\pi\sqrt{2}$ leads
to solutions with $E=1$.  Classical over-barrier solutions of the
original problem with $E > E_S\equiv 1$ are obtained in the limit
$\epsilon \to0$ provided that $T$ also tends to zero while $\tau =
T/\epsilon$ is kept finite.  Different energies correspond to
different values of $\tau$.  And that is what one expects---classical
over--barrier transitions are described by the solutions on the
contour with $T\equiv0$.

Let us now discuss the behavior of the branch points $t_*^\pm$ in complex time
plane at small but finite $\epsilon$ and varying energy. A particularly interesting question
is whether or not the singularities cross our time contour ABCD.
The positions of the singularities in the regularized
problem are  (we set $\Real t_0 = 0$ here, 
see discussion after Eq.~\eqref{Imt0})
\begin{multline*}
  t_{*}^{\pm} = \frac{1}{\sqrt{2E}}\Bigg\{
    in\pi+i\frac{\pi}{2}+i\arg\left(\e^{-i\epsilon /2}
\pm\sqrt{E}\right)
  \\
    \pm\ln\left|\frac{\sqrt{E}+\e^{-i\epsilon /2}}{\sqrt{\e^{-i\epsilon }-
E}}\right|
  \Bigg\}
  \;.
\end{multline*}
The behavior of the real parts of the ``right'' ($t_{*}^+$) and
``left'' ($t_{*}^-$) singularities is simple---as the energy
approaches the barrier height, the singularities move towards large
times, reaching a maximum distance from the origin equal to
$\ln(2/\epsilon)/(2\sqrt{2})$ at $E=1$, and then move back.  This
again shows that our regularization makes the solutions continuous in
energy.  The behavior of the imaginary parts of the positions of the
``right'' singularities, $\Imag( t_{*}^+)$, is also simple: they scale
with energy like $\pi/2\sqrt{2E}$ and stay essentially constant as
energy crosses the value $E = 1$.  The imaginary parts of the
positions of the ``left'' singularities, on the other hand, rapidly
move down near $E = 1$, together with the contour height $T/2$
(Fig.~\ref{fig4}).  So, the singularities do not cross the contour.

Thus, by introducing small parameter $\epsilon $, Eq.~(\ref{Uintnew1D}),
we were able to regularize the system in such a way that the transition from
classically forbidden region to the allowed one is rapid but smooth,
and in the limit $\epsilon \to0$ only the physically relevant
branches of solutions are selected.

Although we have already studied all physically interesting solutions in
non-regularized problem, let us analyze what happens in the limit
$\epsilon\to0$ to the regularized solutions with $0<T<\pi \sqrt{2}$ and $E = 1$.
At this point it is convenient to make the choice~\eqref{Ret0}. Then in the 
limit $\epsilon\to 0$ one obtains a family of solutions with intermediate $T$,
\begin{equation}
  \sinh X = - \e^{-\sqrt{2}\left(t-i\frac{T}{2}+c\right)} \;.
\label{TunOnSph}
\end{equation}
These solutions are no longer real at any finite time, but they
are still real asymptotically, as $t\to+\infty$, where they approach the 
sphaleron, $X = 0$.  For all these
solutions the suppression exponent \eqref{F1dim} is zero.

\begin{figure}
\includegraphics[width=\columnwidth]{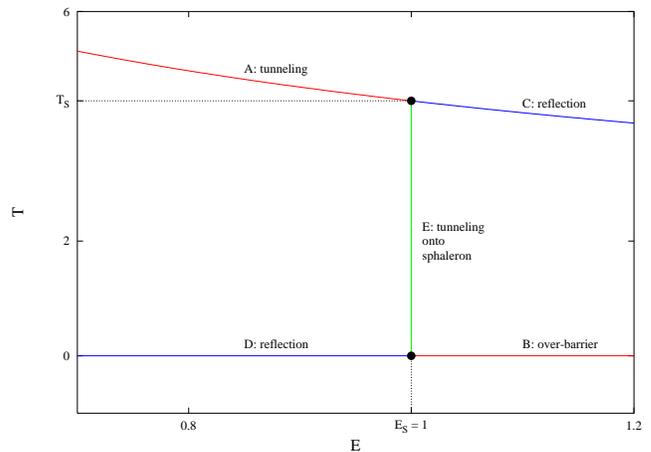}
\caption{The dependence of the period $T$ on energy $E$. Five branches of  
solutions are shown. A: represents
tunneling solutions, Eq.~\eqref{TE}, in the forbidden region 
($E < V_0\equiv 1$);  B: classically allowed transitions,
$T = 0$; (C and D):  unphysical solutions undergoing reflection, 
in allowed and forbidden regions respectively; E  new  
solutions~\eqref{TunOnSph}.}
\label{fig6}
\end{figure}

Thus there are three different branches of solutions merging at each
of the bifurcation points $E=1$, $T/2=0$ and $E = 1$, $T/2=\pi/\sqrt{2}$.
This is shown in Fig.~\ref{fig6}.  
We will see that a somewhat similar
situation occurs in the case of multiple degrees of freedom, but there
the branch analogous to the branch~\eqref{TunOnSph}  is not degenerate in
energy and is precisely the branch of relevant solutions in a certain 
region of parameters.  The generalization of our regularization
technique will enable us to select the physically relevant
branches of solutions.

\section{\label{sec:2D}Quantum mechanics of two degrees of freedom}

The situation we discuss in this and the following section is a transition 
through a potential barrier of an oscillator whose spacing between the levels
is much smaller than the barrier height. This model was already described 
in the Introduction, see \eqref{Hold} for the Hamiltonian. 

As in the one dimensional case, we choose  units with 
$\hbar = 1,$ $m = 1$.  The oscillator frequency $\omega$ is 
of order $1$.  The system is  semiclassical, i.e.\  conditions
\eqref{conditions} are satisfied, if one chooses 
$\sigma = 1/\sqrt{2 \lambda}$, $V_0 =1/\lambda$, where $\lambda$ 
is a small parameter.
Note that $\lambda$ plays a role similar to $\hbar$, determining the magnitude
of quantum fluctuations. At the classical level, this parameter is irrelevant:
after rescaling the variables\footnote{To keep notations simple, we use the 
same symbols $x_1,\;x_2$ for the rescaled variables.}
$x_1 \to x_1/\sqrt{\lambda},\;\; x_2 \to  x_2/\sqrt{\lambda},$
the small parameter enters only through the overall multiplicative factor
$1/\lambda$ in the Hamiltonian. 
Therefore, the semiclassical technique can
be developed as an asymptotic expansion in $\lambda$. 

The properties of the system described by the above Hamiltonian are
made clearer by replacing the variables $x_1,\;x_2$ with the
center-of-mass coordinate $ X \equiv (x_1+x_2)/\sqrt{2}$ and the
relative oscillator coordinate $y\equiv (x_1 - x_2)/\sqrt2$. 
In terms of the latter variables, the Hamiltonian takes the form
\begin{equation}
  H = \frac{p_X^2}{2} + \frac{p_y^2}{2} + \frac{\omega^2}{2} y^2 +
  \frac1\lambda \mathrm{e}^{- \lambda (X+y)^2/2}.
  \label{H}
\end{equation}
The interaction potential
\begin{equation*}
  V_{\mathrm{int}} \equiv \frac{1}{\lambda} \mathrm{e}^{- \lambda (X+y)^2/2}
\end{equation*}
vanishes in the asymptotic regions $X \to \pm\infty$ and describes a 
potential barrier between these regions. At $X\to \pm\infty$ the 
Hamiltonian~\eqref{H} corresponds to an 
oscillator of frequency $\omega$ moving along the center-of-mass coordinate 
$X$. The oscillator asymptotic state may be characterized by 
its excitation number $N$  and total energy 
$ E = \frac{p_X^2}{2} + \omega(N + 1/2)$. 
We are interested in the transmissions through the potential barrier of the 
oscillator with given initial values of $E$ and $N$.

\subsection{\label{sec:Ttheta}$T/\theta$ boundary value problem}

The probability of tunneling from a
state with fixed initial energy $E$ and oscillator excitation number $N$
from the asymptotic region $X \to -\infty$ to any state in the other
asymptotic region $X\to+\infty$ takes the following form:
\begin{equation}\label{TT}
  \mathcal{T}(E,N) = \lim_{t_f - t_i \to \infty}\sum_{f} 
  \left|\bra{f} \e^{- i \hat{H}(t_f-t_i)} \ket{E,N}\right|^2
  \;,
\end{equation}
where it is implicit
that the initial and final states have support only well outside 
the range of the potential, with $X<0$ and $X>0$, respectively.
Semiclassical methods are applicable when the initial energy and excitation 
number are parametrically large,
$$
  E = \tilde{E}/\lambda
  \;,\quad
  N = \tilde{N}/\lambda
  \;,
$$
where $\tilde{E}$ and $\tilde{N}$ are held constant as $\lambda\to 0$.
The transition probability has the exponential form
\begin{equation*}
{\cal T} = D \mathrm{e}^{-\frac{1}{\lambda} F(\tilde{E},\;\tilde{N})}\;,
\end{equation*}
where $D$ is a pre-exponential factor, which we do not consider in
this paper.  Our purpose is to calculate the leading semiclassical
exponent $F(\tilde{E},\; \tilde{N})$. The exponent for tunneling from
the oscillator ground
state~\cite{Rubakov:1992fb,Rubakov:1992ec,Bonini:1999fc} may be
obtained by taking $F(\tilde{E},\tilde{N})$ in the limit
${\tilde{N}\to 0}$.

The exponent $F(\tilde{E},\tilde{N})$ is again related to solutions of
the complexified classical boundary value problem. The derivation is
similar to that given in Sec.~\ref{sec:1D.1}; we present the details
in App.~\ref{app:A}.  The outcome is as follows. In addition to the
Lagrange multiplier $T$ one now has another Lagrange multiplier
$\theta$ which is related to the new parameter $N$ characterizing the
incoming state. We again rescale the variables, $X\to
X/\sqrt{\lambda}$, $y\to y/\sqrt{\lambda}$, and omit tilde over the
rescaled quantities $\tilde{E},\;\tilde{N}$.  As above, the boundary
value problem is conveniently formulated on the contour ABCD in the
complex time plane (see Fig.~\ref{fig2}), with the imaginary part of
the initial time equal to $T/2$.  The coordinates $X(t),\; y(t)$ must
satisfy the complexified equations of motion on the internal points of
the contour, and be real in the asymptotic future (region D):
\begin{subequations}\label{bc}
\begin{eqnarray}
  && \frac{\delta S}{\delta X(t)}=\frac{\delta S}{\delta y(t)} = 0
  \;,\\[1ex]
  &&
  \begin{array}{l}
    \displaystyle \Imag y(t) \to 0, \\[1ex]
    \displaystyle \Imag X(t) \to 0,  
  \end{array}
  \qquad\text{as } t \to +\infty \;.
  \label{bcplus}
\end{eqnarray}
In the asymptotic past (region A of the contour, where $t = t' + i
T/2$, $t'$ is real negative) one can neglect the interaction potential
$U_\mathrm{int}$, and the oscillator decouples:
\begin{equation*}
  y = \frac{1}{\sqrt{2 \omega}}(u \mathrm{e}^{- i \omega t'} +  
  v \mathrm{e}^{i \omega t'}).
\end{equation*}
The boundary conditions in the asymptotic past, $t' \to -\infty$, are
that the center-of-mass coordinate $X$ must be real, while the complex
amplitudes of the decoupled oscillator must be linearly related,
\begin{equation}\label{bcminus}
\begin{array}{l}
\Imag X \to 0,\\[1ex]
  v \to \e^\theta u^*,\; 
\end{array}
  \qquad  \text{as } t' \to -\infty\;.
\end{equation}
\end{subequations}
The boundary conditions \eqref{bcplus} and \eqref{bcminus} make, in
fact, eight real conditions (since, e.g.,\ $\Imag X(t')\to0$ implies
that both $\Imag X$ and $\Imag\dot{X}$ tend to zero), and completely
determine a solution, up to time translation invariance (see
discussion in App.~\ref{app:A}).

It is shown in App.~\ref{app:A} that a solution to this boundary value
problem is, in fact, an extremum of the functional (cf.~\eqref{F1dim})
\begin{multline}\label{F}
  F[X,y;\;X^*,y^*;\;T,\theta] = - iS[X,y] + i S[X^*,y^*] \\
    - ET - N\theta
    + \text{Boundary terms}\;.
\end{multline}
The value of this functional at the extremum gives the exponent for the
 transition
probability
\begin{equation}\label{Fshort}
  F(E,\;N) = 2 \Imag  S_0(T,\;\theta) - E T - N \theta\;,
\end{equation}
where $S_0$ is the action of the solution (integrated by parts, see
App.~\ref{app:A}),
\begin{equation*}
  S_0 = \int \! dt\left(\!
    - \frac12 X\frac{d^2 X}{dt^2} - \frac12 y \frac{d^2 y}{dt^2} - 
    \frac12 \omega^2 y^2 - U_{\mathrm{int}}(X,y)
  \right).
\end{equation*}
The values of the Lagrange multipliers $T$ and $\theta$ are related 
to energy and excitation number as follows,
\begin{eqnarray}
  E(T,\theta) &=& \frac\partial{\partial T} 2 \Imag  S_0(T,\theta)\;,
  \label{E}\\
  N(T,\theta) &=& \frac\partial{\partial \theta} 2 \Imag  S_0(T,\theta)\;.
  \label{N}
\end{eqnarray}
Making use of Eq.~\eqref{Fshort}, it is straightforward to check also the
inverse Legendre transformation formulas,
\begin{eqnarray}
  T(E,\;N) &=& - \frac{\partial}{\partial E} F(E,\;N) \;,
  \label{T}\\
  \theta(E,\;N) &=& - \frac{\partial}{\partial N} F(E,\;N) \;.
  \label{theta}
\end{eqnarray} 
One can also check that the right hand side of Eq.~\eqref{E} coincides with
the energy of the classical solution, while the right hand side of
Eq.~\eqref{N} is equal to the classical counterpart of the occupation
number,
\begin{equation}\label{energyeq}
  E=\frac{\dot{X}^2}{2}+\omega N
  \;;\qquad
  N = uv\;.
\end{equation}
So, one may either search for the  values of $T$ and
$\theta$ that correspond to some given $E$ and $N$, or, following a
 computationally simpler procedure, solve the boundary
value problem \eqref{bc} for given $T$ and $\theta$ and then find the
corresponding values of $E$ and $N$ from
Eq.~\eqref{energyeq}.

Let us discuss some subtle points of the boundary value problem
\eqref{bc}. First, one notices that the condition of asymptotic
reality \eqref{bcplus} does not always coincide with the condition of
reality at finite time.  Of course, if the solution approaches the
asymptotic region $X\to+\infty$ on the part CD of the contour, the
asymptotic reality condition \eqref{bcplus} implies that the solution
is real at any \emph{finite} positive $t$.  Indeed, the oscillator
decouples as $X\to+\infty$, so the condition \eqref{bcplus} means that
its phase and amplitude, as well as $X(t)$, are real as $t\to
+\infty$.  Due to equations of motion, $X(t)$ and $y(t)$ are real on
the entire CD--part of the contour.  This situation corresponds to the
transition directly to the asymptotic region $X\to+\infty$.  However,
the situation can be drastically different if the solution on the
final part of the time contour remains in the interaction region.  For
example, let us imagine that the solution approaches the saddle point
of the potential $X = 0$, $y = 0$ as $t\to+\infty$.  Since one of the
excitations about this point is unstable, there may exist solutions
which approach this point {\it exponentially} along the unstable
direction, i.e.\ $X(t),\; y(t)\propto \mathrm{e}^{- \mathrm{const}
\cdot t}$ with possibly complex pre-factors.  In this case the
solution may be complex at any finite time, and become real only
asymptotically, as $t\to+\infty$.  Such solution corresponds to
tunneling to the saddle point of the barrier, after which the system
rolls down classically towards $X\to+\infty$ (with probability of
order $1$, inessential for the tunneling exponent $F$).  We will see
in Sec.~\ref{sec:reg_forb} that the situation of this sort indeed
takes place for some values of energy and excitation number.

Second, since at large negative time (in the asymptotic region
$X\to-\infty$) the interaction potential disappears, it is
straightforward to continue the asymptotics of the solution to the
real time axis.  For solutions satisfying \eqref{bcminus} this gives
at large negative time
\begin{eqnarray}
&&  y(t) = \frac{1}{\sqrt{2\omega}}\left(
    u \e^{-\frac{\omega T}{2}}\e^{-i\omega t} +  
    u^*\e^{\theta+\frac{\omega T}{2}}\e^{i\omega t}
  \right) \;,\nonumber
  \\
&&  \Imag X(t) = -i\frac{T}{2}p_X \;.\nonumber
\end{eqnarray}
We see that the dynamical coordinates on the negative side of the real
time axis are generally complex.  For solutions approaching the
asymptotic region $X\to+\infty$ as $t\to +\infty$ (so that $X$ and $y$
are exactly real at finite $t>0$), this means, in complete analogy to
the one-dimensional case, that there should exist a branch point in
the complex time plane: the contour in Fig.~\ref{fig2} winds around
this point and cannot be deformed to the real time axis.  This
argument \emph{does not} work for solutions ending in the interaction
region as $t\to+\infty$, so branch points between the AB--part of the
contour and the real time axis may be absent.  We will see in
Sec.~\ref{sec:reg_forb} that this is indeed the case in our model for
a certain range of $E$ and $N$.

\subsection{\label{sec:over-barrier}Over-barrier transitions: boundary
of the classically allowed region $E_0(N)$}

Before studying the exponentially suppressed transitions, let us
consider the classically allowed ones. To this end, let us study the
classical evolution, in which the system is initially located at large
negative $X$ and moves with positive center-of-mass velocity towards
the asymptotic region $X \to +\infty$.  The classical dynamics of the
system is specified by four initial parameters.  One of them (e.g.,
the initial center-of-mass coordinate) fixes the invariance under time
translation, while the other three are the total energy $E$, the
initial excitation number of the $y$--oscillator, defined in classical
theory as $N\equiv E_\mathrm{osc}/\omega$, and the initial oscillator
phase $\varphi_i$.

The classically allowed region is the region of the $E$--$N$ plane where
over--barrier transitions are possible for some value(s) of the initial
oscillator phase\footnote{Note that the corresponding classical solutions 
obey the boundary conditions~\eqref{bcplus}, \eqref{bcminus} with 
$T = \theta = 0$, i.e., they are solutions to the boundary value 
problem~\eqref{bc}.}.  For given $N$, at large enough $E$
the system can certainly evolve to the other side of the barrier.  On the other
hand, if $E$ is smaller than the barrier height, the system
definitely undergoes reflection.  Thus, there exists some boundary
energy $E_0(N)$, such that classical
transitions are possible for $E > E_0(N)$, while for
$E < E_0(N)$ they do not occur for \emph{any} initial phase
$\varphi_i$.  The line $E_0(N)$ represents the boundary of
the classically allowed region.
We have calculated $E_0(N)$ numerically for $\omega = 0.5$: the result is 
shown in Fig.~\ref{fig1}
and reproduced in Fig.~\ref{fig7}

\begin{figure}
  \begin{center}
    \includegraphics[width=\columnwidth]{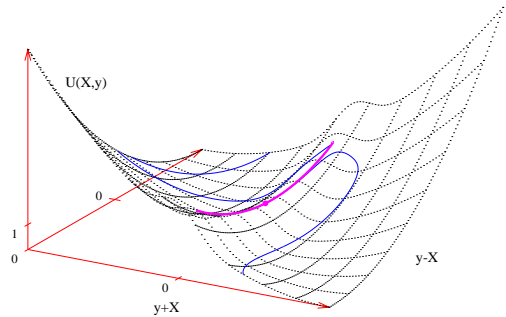}
  \end{center}
  \caption{The potential (dotted lines)
    in the vicinity of the sphaleron ${(X = 0,\;y = 0)}$ (marked by
    the point), excited sphaleron (thick line)
    corresponding to the point $(E,N) = (1.985,3.72)$
    at the boundary of the classically allowed region, and the
    trajectory of the solution which is close to this excited
    sphaleron (thin line).  In this figure the asymptotic regions
    $X\to\pm\infty$ are along the diagonal.}
  \label{fig8} 
\end{figure}

An important point of the boundary $E_0(N)$ corresponds to the static
unstable classical solution $X(t) = y(t) = 0$, the sphaleron (cf.\
Refs.~\cite{Manton:1983nd,Klinkhamer:1984di}).  It is the saddle point
of the potential $U(X,y)\equiv {\omega^2
y^2}/{2}+U_{\mathrm{int}}(X,y)$ and has exactly one unstable
direction, the negative mode (see Fig.~\ref{fig8}).  The sphaleron
energy $E_\mathrm{S} = U(0,0) = 1$ determines the minimum value of the
function $E_0(N)$.  Indeed, while classical over--barrier transitions
with ${E < E_{\mathrm{S}}}$ are impossible, the over--barrier solution
with slightly higher energy can be obtained as follows: at the point
$X=y=0$, one adds momentum along the negative mode, thus ``pushing''
the system towards $X\to+\infty$.  Continuing this solution backwards
in time one finds that the system tends to $X\to-\infty$ for large
negative time and has a certain oscillator excitation number.
Solutions with energy closer to the sphaleron energy correspond to
smaller ``push'' and thus spend longer time near the sphaleron.  In
the limiting case when the energy is equal to $E_{\mathrm{S}}$, the
solution spends an infinite time in the vicinity of the sphaleron.
This limiting case has a definite initial excitation number
$N_{\mathrm{S}}$, so that $E_0(N_{\mathrm{S}}) = E_\mathrm{S}$ (see
Fig.~\ref{fig7}).  The value of $N_\mathrm{S}$ is unique because there
is exactly one negative direction of the potential in the vicinity of
the sphaleron.

In complete analogy to the features of the over--barrier classical
solutions near the sphaleron point ($E_\mathrm{S}$, $N_{\mathrm{S}}$),
one expects that as the values of $E,\;N$ approach any other boundary
point $(E_0(N),\;N)$, the corresponding over--barrier solutions will
spend more and more time in the interaction region, where
$U_{\mathrm{int}} \ne 0$.  This again follows from a continuity
argument.  Namely, let us first fix the initial and final times, $t_i$
and $t_f$.  If in this time interval a solution with energy $E_1$
evolves to the other side of the barrier and a solution with energy
$E_2$ and the same oscillator excitation number is reflected back,
there exists an intermediate energy at which the solution ends up at
$t = t_f$ in the interaction region.  Taking the limit $t_f \to
+\infty$ and $(E_1-E_2)\to0$, we obtain a point at the boundary of the
classically allowed region and a solution tending asymptotically to
some unstable time dependent solution that spends infinite time in the
interaction region.  We call the latter solution \emph{excited
sphaleron;} it describes some (in general nonlinear) oscillations
above the sphaleron along the stable direction in coordinate space.
Therefore, every point of the boundary ${(E_0(N),\;N)}$ corresponds to
some excited sphaleron.

We display in Fig.~\ref{fig8} a solution, found numerically in our
model, that tends to an excited sphaleron.  We see that this
trajectory is, roughly speaking, orthogonal to the unstable direction
at the saddle point $({X = 0,}\;{y = 0})$.

\subsection{\label{sec:bifurcation}Suppressed transitions: bifurcation
    line $E_1(N)$}

\begin{figure}
  \centerline{\includegraphics[width=\columnwidth]{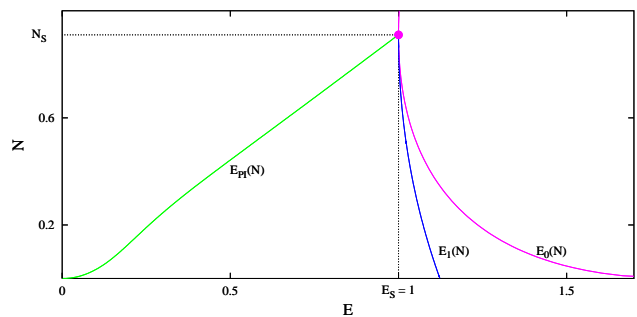}}
  \caption{The boundary of the classically allowed region 
    $E_0(N)$, the bifurcation line $E_1(N)$ and the
    line of the periodic instantons $E_{PI}(N)$.}
  \label{fig7} 
\end{figure}

Let us now turn to classically forbidden transitions, and consider the 
boundary value problem~\eqref{bc}. It is relatively straightforward to 
obtain numerically  solutions for $\theta = 0$.
The boundary conditions
\eqref{bcplus}, \eqref{bcminus} in this case take the form of 
reality conditions in the asymptotic future and past. It can be shown
(see Ref.~\cite{Khlebnikov:1991th}) that the physically relevant solutions
with $\theta = 0$ are real on the entire contour ABCD of Fig.~\ref{fig2} and 
describe nonlinear oscillations in the
upside-down potential on the Euclidean part BC of the contour. The period of 
the
oscillations is equal to $T$, so that the points B and C are two different 
turning
points, where $\dot{X} = \dot{y} = 0$.  These real Euclidean
solutions are called periodic instantons
(see Ref.~\cite{Khlebnikov:1991th}). A practical technique for obtaining these 
solutions numerically on the Euclidean part BC consists in minimizing
 the Euclidean action (for example with the method of 
conjugate gradients, see Ref.~\cite{Bonini:1999cn} for details). 
The solutions on the entire contour are then obtained by solving numerically 
the Cauchy problem, forward in time along the line CD
and backward in time along the line BA. Having the solution in asymptotic past
(region A), one then calculates its energy and excitation
number~\eqref{energyeq}. The solutions to this Cauchy problem are obviously real,
so the boundary conditions~\eqref{bcplus}, \eqref{bcminus} are indeed 
satisfied with $\theta = 0$. It is worth noting that solutions with $\theta = 0$
are similar to the ones in quantum mechanics of one degree of freedom, see 
Fig.~\ref{fig3}.
The line of 
periodic instantons in $E$--$N$ plane in our model is
shown in Fig.~\ref{fig7}, again for $\omega = 0.5$.

Once the solutions with $\theta = 0$ are found, it is natural to try to
cover the entire classically forbidden region of the $E-N$ plane with
a deformation procedure,
by moving in small steps in $\theta$ (and $T$).  The solution to the boundary 
value problem with $(T + \Delta T,\; \theta + \Delta \theta)$ may be
obtained numerically, by applying an iteration technique, with the known
solution at $(T,\;\theta)$ serving as the zeroth-order approximation\footnote{
In practice, the Newton--Raphson method is particularly convenient (see 
Refs.~\cite{Bonini:1999cn},~\cite{Kuznetsov:1997az},~\cite{Bezrukov:2001dg}).}.
Provided the solutions have correct ``topology'' at each step, 
i.e.\ $X\to -\infty$ and $X\to +\infty$ on parts A and D of the contour, the 
solutions obtained by this ``walking'' procedure are physically relevant. 
However, the ``walking'' method fails to produce relevant solution  if there are 
bifurcation points, where the
physical branch of solutions  merges to an unphysical branch. As there are 
unphysical solutions close to physical ones in the vicinity of a
bifurcation point, one  cannot ``walk''
near these points.

\begin{figure}
  \begin{center}
    \includegraphics[width=\columnwidth]{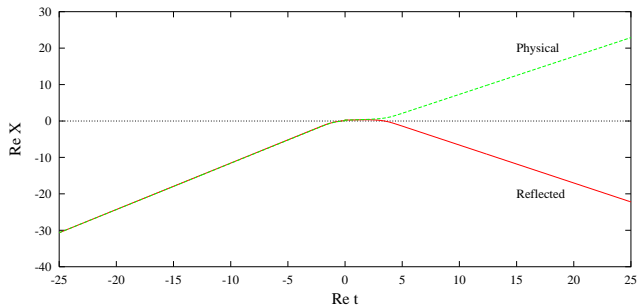}
  \end{center}
  \caption{The dependence of the tunneling coordinate $X$ on time for 
     two solutions with nearly the same energy and initial excitation number. 
	The physical solution tunnels to the asymptotic region $X\to +\infty$,
while the unphysical one gets reflected back to $X\to -\infty$.
      The physical solution has $E = 1.028,\;N =
    0.44$, while the unphysical one has $E = 1.034,\;N =
    0.44$. These two solutions are close to the point at the bifurcation
line $E_1(N = 0.44) = $ 1.031.}
  \label{fig9}
\end{figure}

In our model, the walking method produces correct solutions to the
$T/\theta$ boundary value problem in a large region of the $E-N$ plane, 
where $E < E_1(N)$.  However, at
sufficiently high energy $E > E_1(N)$, where
$E_1(N) \gtrsim E_\mathrm{S}$,  the deformation procedure 
generates solutions that bounce back from the barrier (see
Fig.~\ref{fig9}), i.e.\ have the wrong ``topology''. 
Clearly, the latter do not describe the tunneling transitions of
interest.  Therefore, if the semiclassical method is applicable at all
in the region $E_1(N) < E < E_0(N)$, there exists another, physical
branch of solutions.  In that case the line $E_1(N)$ is the
bifurcation line, where the physical solutions ``meet'' the ones with
wrong ``topology''.  Walking in small steps in $\theta$ and $T$ is
useless in the vicinity of this bifurcation line, and one needs to
introduce some trick to find the relevant solutions beyond that
line. The bifurcation line $E_1(N)$ for our quantum mechanical problem
of two degrees of freedom with $\omega = 0.5$ is shown in
Fig.~\ref{fig7}.

The loss of topology beyond a certain bifurcation line in $E-N$ plane
is by no means a property of our model only.  This phenomenon has been
observed in field theory models, in the context of both induced false
vacuum decay~\cite{Kuznetsov:1997az} and baryon-number violating
transitions in gauge theory~\cite{Bezrukov:2001dg} (in field-theoretic
models, the parameter $N$ is the number of incoming particles).  In
all cases, the loss of topology prevented one to compute the
semiclassical exponent for the transition probability in an
interesting region of relatively high energies, where the suppression
of tunneling may be fairly weak.

Coming back to the quantum mechanics of two degrees of freedom, we
point out that the properties of tunneling solutions approaching the
bifurcation line $E_1(N)$ from the left, are in some sense similar to
the properties of tunneling solutions in one-dimensional quantum
mechanics near the boundary of the classically allowed region. Again
by continuity, these solutions of our two-dimensional model spend a
long time in the interaction region; this time tends to infinity on
the line $E_1(N)$. Hence, on any point of this line, there is a
solution that starts in the asymptotic region left of the barrier, and
ends up on an excited sphaleron. We already mentioned in
Sec.~\ref{sec:Ttheta} that such a behavior is indeed possible because
of the existence of an unstable direction near the (excited)
sphaleron, even for complex initial data.  We have suggested in
Sec.~\ref{sec:1D.2} a trick to deal with this situation---this is our
regularization technique.

\section{\label{sec:4}Regularization technique}

In this Section we further develop our regularization technique, and
find the physically relevant solutions between the lines $E_1(N)$ and
$E_0(N)$.  We will see that all solutions from the new branch (and not
only on the lines $E_0(N)$ and $E_1(N)$) correspond to tunneling onto
the excited sphaleron.  These solutions would be very difficult, if at
all possible, to obtain directly, by solving numerically the
non-regularized classical boundary value problem~\eqref{bc}: they are
complex at finite times, and become real only asymptotically as $t\to
+\infty$, whereas numerical methods require working with finite time
intervals.

As an additional advantage, our
regularization technique enables one to obtain a family of the over--barrier
solutions, that covers all the classically allowed region, including its 
boundary. This may be of interest in models with large number of degrees of 
freedom and in field theory, where finding the boundary of the classically
allowed region by direct methods is difficult (see e.g., 
Ref.~\cite{Rebbi:1995zw} for discussion of this point).

\subsection{\label{sec:reg_forb}Regularized problem in the classically
forbidden region}

The main idea of our method  is
to ``regularize'' the equations of motion by adding a term
proportional to a small parameter $\epsilon$ so that configurations staying 
for an infinite time near the sphaleron no
longer exist among the solutions of the $T/\theta$ boundary value
problem.  After performing the regularization we explore all the
classically forbidden region without crossing the bifurcation line.
Taking then the limit ${\epsilon \to 0}$ we reconstruct the correct
values of $F$, $E$ and $N$.

When formulating the regularization technique it is more convenient to
work with the functional $F[X,y;X^*,y^*;T,\theta]$ itself rather than
with the equations of motion.  We slightly modify the form of $F$,
Eq.~\eqref{F}, so that $F$ is no longer extremized by configurations
approaching the excited sphalerons asymptotically.  To achieve this,
we add to the original functional~\eqref{F} a new term of the form $2
\epsilon T_{\mathrm{int}}$, where $T_{\mathrm{int}}$ estimates the
time the solution ``spends'' in the interaction region.  The parameter
of regularization $\epsilon$ is the smallest one in the problem, so
any regular extremum of the functional $F$ (the solution that spends
finite time in the region $U_{\mathrm{int}} \ne 0$) changes only
slightly after the regularization.  At the same time, the excited
sphaleron configuration has $T_{\mathrm{int}} = \infty$ which leads
infinite value of the regularized functional $F_{\epsilon} \equiv F +
2 \epsilon T_\mathrm{int}$.  Hence, the excited sphalerons are not
stationary points of the regularized functional.

For the problem at hand, $U_{\mathrm{int}} \sim 1$ in the interaction 
region, and $T_\mathrm{int}$ can be defined as follows,
\begin{equation}
T_{\mathrm{int}} = \frac12 \left[\int dt\; U_{\mathrm{int}}(X,y) + 
\int dt\; U_{\mathrm{int}} (X^*,y^*)\right] \;.\label{Tint}
\end{equation}
We notice that $T_\mathrm{int}$ is real, and that the regularization 
is equivalent to the 
multiplication of the interaction potential by a complex factor
\begin{equation}
U_{\mathrm{int}}\to (1 - i \epsilon) U_{\mathrm{int}} = 
\mathrm{e}^{- i \epsilon} U_{\mathrm{int}}  + O(\epsilon^2)
\;.\label{Uintnew}
\end{equation}
This results in the corresponding change of the classical equations of
motion, while the boundary conditions~\eqref{bcplus},~\eqref{bcminus} 
remain unaltered.

We still have
to understand whether solutions with $\epsilon \ne 0$ exist at
all.  The reason for the existence of such solutions is as follows.
Let us consider a well-defined (for $\epsilon > 0$) matrix element
\begin{equation*}
  \mathcal{T}_{\epsilon} = \lim\limits_{t_f - t_i \to \infty}
  \sum\limits_{f} \left|\langle f|
\mathrm{e}^{ (- i \hat H - \epsilon U_\mathrm{int})(t_f - t_i)} 
| E,N \rangle\right|^2 \;,
\end{equation*}
where $|E,\;N\rangle$ denotes, as before, the incoming state with given energy 
and number of particles.  The quantity ${\cal T}_{\epsilon}$ has a well 
defined limit as $\epsilon \to 0$, equal to 
the tunneling probability~\eqref{TT}. As the saddle point of the regularized
functional $F_{\epsilon} \equiv F + 2 \epsilon T_{\mathrm{int}}$ gives
the semiclassical exponent for the quantity ${\cal
  T}_{\epsilon}$, one expects that such saddle point indeed exists.

Therefore, the regularized $T/\theta$ boundary value problem is
expected to have solutions necessarily spending finite time in the
interaction region. By continuity, these solutions do not experience 
reflection from the barrier,
if one makes use of the ``walking'' procedure starting from solutions with correct ``topology''.
The line $E_1(N)$ is no longer a bifurcation line of the regularized 
system, so the walking procedure enables one  to cover the entire forbidden region.
The semiclassical  suppression factor of the original problem is recovered 
in the limit $\epsilon \to 0$.

It is worth noting that the interaction time is Legendre
conjugate to $\epsilon$,
\begin{equation}\label{TintThetaE}
  T_{\mathrm{int}} =  \frac{1}{2}
  \frac{\partial}{\partial \epsilon} F_{\epsilon}(E,N,\epsilon)
  \;.
\end{equation}
Useful equations can also be obtained from Eqs.~\eqref{E} and \eqref{N}:
\begin{eqnarray}
  E &=& E\vert_{\epsilon = 0} + 2\epsilon  
        \frac{\partial T_{\mathrm{int}}}{\partial T}  + O(\epsilon)
  \;,\label{EThetaE}\\
  N &=& N\vert_{\epsilon = 0} + 2\epsilon  
        \frac{\partial T_{\mathrm{int}}}{\partial \theta}  + O(\epsilon)
  \;.\label{NThetaE}
\end{eqnarray}
These equations show how the values of $E$ and $N$ change with
$\epsilon$---away from the would-be bifurcation line $E_1(N)$ the
values of $E$ and $N$ depend slightly on $\epsilon$, while at $E \sim
E_1(N)$ where $T_{\mathrm{int}}$ takes large values (for small
$\epsilon$), the dependence is strong and results in a considerable
shift of the point $(E,N)$.  Eqs.~(\ref{TintThetaE}),
(\ref{EThetaE}), (\ref{NThetaE}) may be used as a check of numerical
calculations.

We implemented the regularization procedure numerically. 
To solve the boundary value problem, we make use of 
the computational methods described in Ref.~\cite{Bonini:1999fc}. In our calculations
we used $\omega = 0.5$.
To obtain the semiclassical tunneling exponent in the region between
the bifurcation line $E_1(N)$ and boundary of the allowed region, $E_0(N)$,
we began with a solution to the non-regularized problem deep in the forbidden 
region (i.e., at $E < E_1(N)$). For that value of $T$ and $\theta$ we increased 
the value of $\epsilon$ from zero to a certain small positive number.
Then we changed $T$ and $\theta$ in small steps, keeping $\epsilon$ finite,
and found solutions to the regularized problem in the region 
$E_1(N) < E < E_0(N)$. These solutions had correct ``topology'', i.e.\ they
indeed ended up in the asymptotic region $X \to +\infty$. Finally, we 
lowered $\epsilon$ and extrapolated $F$, $E$ and $N$ to the limit 
$\epsilon \to 0$.
\begin{figure}
  \begin{center}
    \includegraphics[width=\columnwidth]{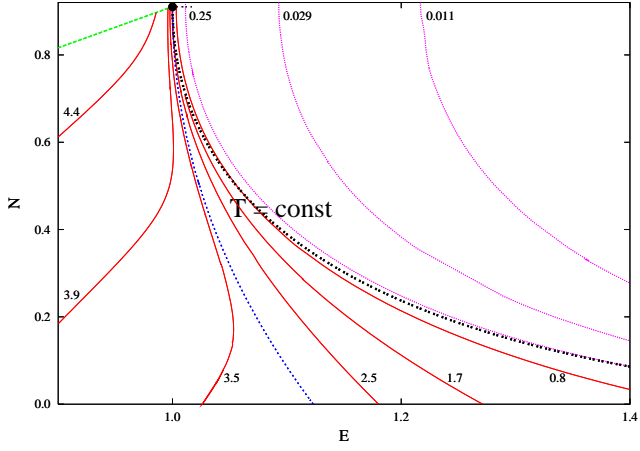}
  \end{center}
\caption{Lines $T = \mathrm{const}$ for $\epsilon = 0.002$.  
Numbers near each line show the values of $T$.
The boundary of the classically allowed region, the bifurcation line and 
the line of periodic instantons are shown by dashed lines.}
\label{fig10}
\end{figure}

The lines $T = \mathrm{const}$ for $\epsilon = 0.002$ are shown in
Fig.~\ref{fig10}.  We see that at $\epsilon \ne 0$ these lines are
smooth and extend from $\theta = 0$ (the line of periodic instantons)
to $\theta = \infty$ (the line $N = 0$).  They indeed cover all the
tunneling region without any irregularities at the bifurcation line
$E_1(N)$.

\begin{figure}
\begin{center}
\includegraphics[width=\columnwidth]{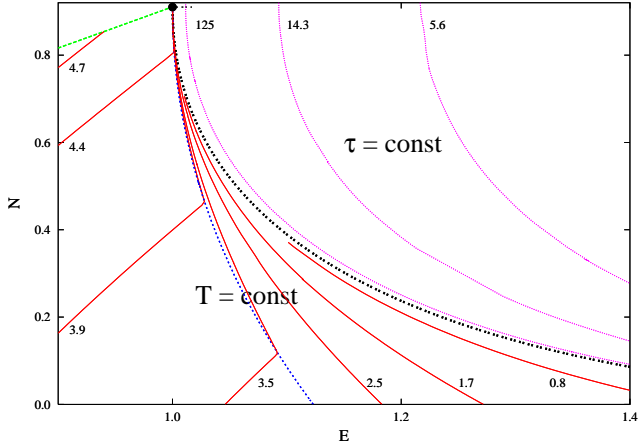}
\end{center}
\caption{ The lines $T = \mathrm{const}$ (in the forbidden region) and
$\tau = \mathrm{const}$ (in the allowed region), see
Sec.~\ref{sec:4.3}, in the limit $\epsilon \to 0$.  Numbers near the
lines are the values of $T$ and $\tau$.  }
\label{fig11}
\end{figure}

\begin{figure}
\begin{center}
\includegraphics[width=\columnwidth]{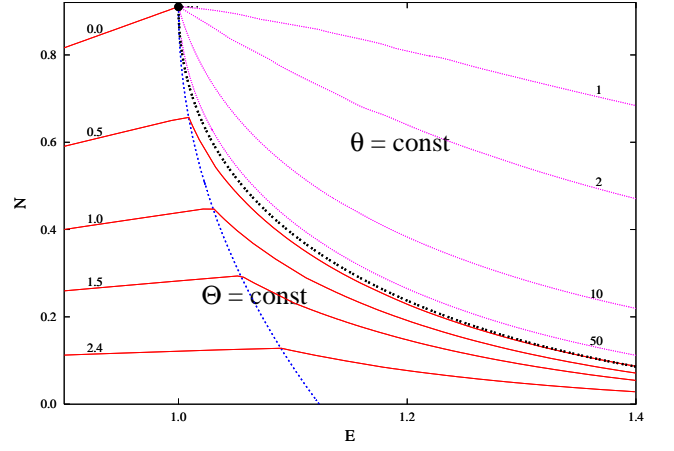}
\end{center}
\caption{The lines $\theta = \mathrm{const}$ and 
$\vartheta = \mathrm{const}$ in the limit $\epsilon \to 0$.
Numbers near the lines are the values of $\theta$ and $\vartheta$.
}
\label{fig12}
\end{figure}

\begin{figure}
\begin{center}
\includegraphics[width=\columnwidth]{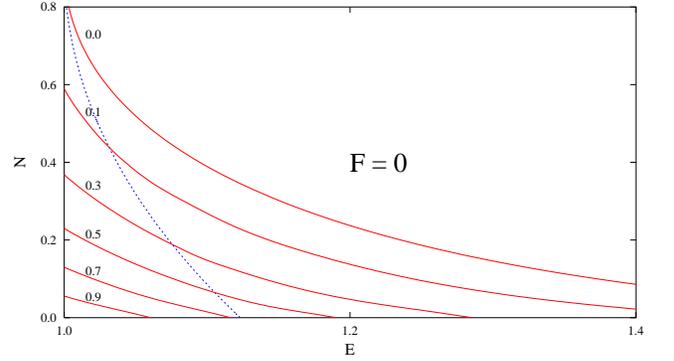}
\end{center}
\caption{The lines $F = \mathrm{const}$ at energies above $E_S = 1$. 
Dashed line shows $E_1(N)$.
}
\label{fig13}
\end{figure}

The same lines $T = \mathrm{const}$, but now in the limit $\epsilon
\to 0$ are shown in Fig.~\ref{fig11}; the lines ${\theta =
\mathrm{const}}$, again in the limit $\epsilon\to 0$, are presented in
Fig.~\ref{fig12}. It is seen that the functions $T(E,N)$ and
$\theta(E,N)$ are continuous, while their first derivatives have
discontinuities at the bifurcation line $E_1(N)$.  As $T$ and $\theta$
are the derivatives of the tunneling exponent $F(E,N)$, see
Eqs.~\eqref{T},~\eqref{theta}, the suppression exponent $F$ itself has
discontinuities of the second derivatives only.  The lines $F =
\mathrm{const}$ are shown in Fig.~\ref{fig13}. We see that the
function $F$ and its first derivatives are indeed continuous.

\begin{figure}
\includegraphics[width=\columnwidth]{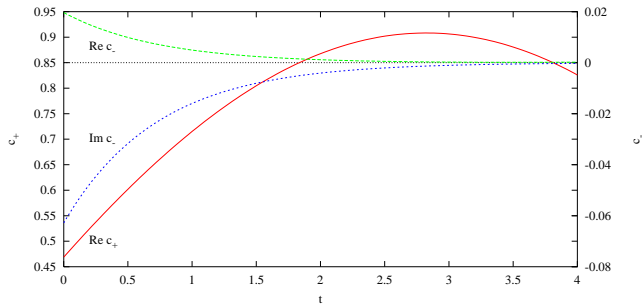}
\caption{The large-time behavior of a solution with 
$\epsilon = 0$ at $(E = 1.05,\; N=0.43)$.
The coordinates $X$ and $y$ are decomposed in a basis of the eigenmodes near 
the sphaleron.  Note that $\Imag c^+ = 0$.}
\label{fig14}
\end{figure}

Let us consider more carefully the solutions in the region $E_1(N) < E
< E_0(N)$ which we obtain in the limit $\epsilon\to0$.  They belong to
a new branch, and thus may exhibit new physical properties.  Indeed,
we found that, as the value of $\epsilon$ decreases to zero, the
solution at {\it any} point $(E,N)$ with $E_1(N) < E < E_0(N)$ spends
more and more time in the interaction region. The limiting solution
corresponding to $\epsilon = 0$ has infinite interaction time: in
other words, it tends, as $t\to +\infty$, to one of the excited
sphalerons.  The resulting physical picture is that at large enough
energy (i.e., at $E > E_1(N)$), the system prefers to tunnel exactly
onto an unstable classical solution, excited sphaleron, that
oscillates about the top of the potential barrier!  To demonstrate
this, we have plotted in Fig.~\ref{fig14} the solution
$\vec{x}(t)\equiv (X(t),y(t))$ at large times, after taking
numerically the limit $\epsilon \to 0$. To understand this figure, one
recalls that the potential near the sphaleron point $X = y = 0$ has
one positive mode and one negative mode.  Namely, by introducing new
coordinates $c_+$, $c_-$,
\begin{eqnarray}
&&X = \cos\alpha \;c_+ + \sin\alpha \; c_-\;,\nonumber\\
&&y = -\sin\alpha \;c_+ + \cos\alpha \; c_-\;,\nonumber\\
&& \mathrm{ctg } 2 \alpha = - \frac{\omega^2}{2}\;,\nonumber
\end{eqnarray}
one writes, in the vicinity of the sphaleron,
\begin{equation*}
H = 1 + \frac{p_+^2}{2} + \frac{p_-^2}{2} + \frac{\omega_+^2}{2} c_+^2 - 
\frac{\omega_-^2}{2} c_-^2\;,
\end{equation*}
where
\begin{equation*}
\omega_\pm = - 1 + \frac{\omega^2}{2} \pm \sqrt{1 + \frac{\omega^4}{4}}\;.
\end{equation*}
Since the solutions to the $T/\theta$ boundary value problem are
complex, the coordinates $c_+$ and $c_-$ are complex too.  We show in
Fig.~\ref{fig14} real and imaginary parts of $c_+$ and $c_-$ at large
real time $t$ (part CD of the contour).  We see that while $\Real c_+$
oscillates, the unstable coordinate $c^-$ approaches asymptotically
the sphaleron value: $c_- \to 0$ as $t \to +\infty$.  The imaginary
part of $c_-$ is non-zero at any finite time.  This is the reason for
the failure of straightforward numerical methods in the region $E >
E_1(N)$: the solutions from the physical branch do not satisfy the
conditions of reality at any large but finite final time.  We have
pointed out in Sec.~\ref{sec:Ttheta} that this can happen only if the
solution ends up near the sphaleron, which has a negative mode.

\begin{figure}
  \begin{center}
    \includegraphics[width=\columnwidth]{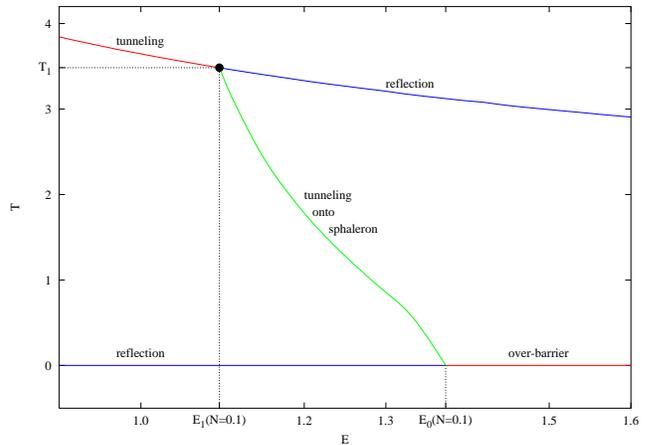}\\
	(a)\\
    \includegraphics[width=\columnwidth]{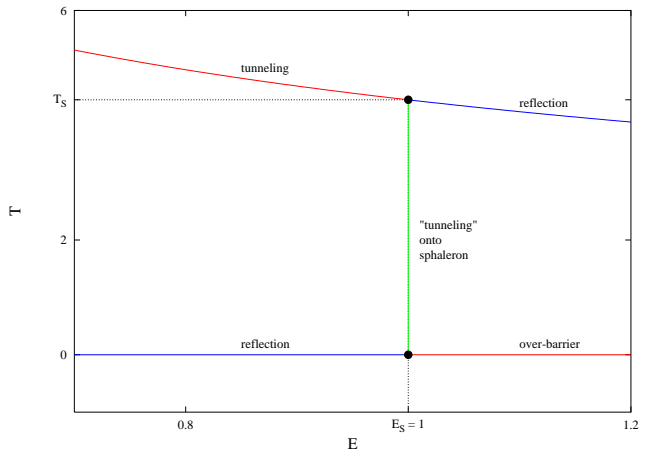}\\
	(b)
  \end{center}
  \caption{Dependence of the parameter $T = - \partial F/\partial E$ on energy 
for (a) the two dimensional model with fixed $N=0.1$ and (b)the one dimensional model.
The lower picture is the same as Fig.~\ref{fig6}. 
Different lines correspond to different branches of classical solutions
to $T/\theta$ boundary value problem. The branches labelled ``reflection''
have wrong ``topology''.}
\label{fig15}
\end{figure}

It is interesting to compare the branches of solutions in the two--
and one--dimensional models (the latter was discussed in
Sec.~\ref{sec:1D}).  The dependence $T(E)$ in these models is shown in
Fig.~\ref{fig15} (for the two dimensional problem the graph
corresponds to constant oscillator excitation number $N=0.1$).  We see
that the structure of the branches is similar, though in the
one--dimensional case the solutions corresponding to ``tunneling''
onto the sphaleron are degenerate in energy and are not really useful
for finding the transition exponent. In the two--dimensional model, on
the other hand, the similar branch is the one relevant at $E_1(N) < E
< E_0(N)$.

\subsection{\label{sec:4.2}Regularization technique versus exact
    quantum--mechanical solution}

\begin{figure}
  \begin{center}
    \includegraphics[width=\columnwidth]{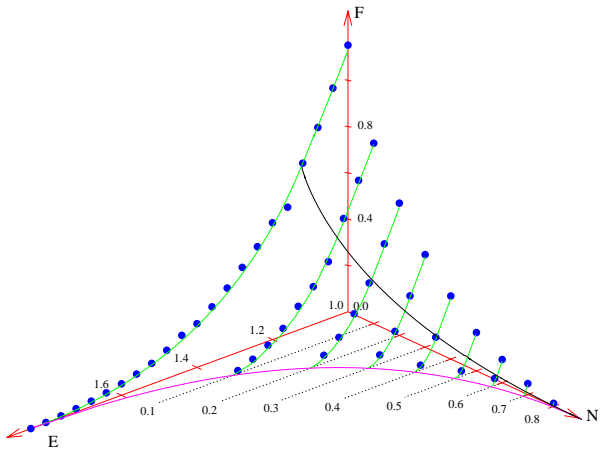}
  \end{center}
  \caption{The tunneling exponent $F(E,N)$
    in the region $E > E_{\mathrm{S}} = 1$.  The lines
    show the semiclassical results while points represent the exact
    ones, obtained by solving the Schr\"odinger equation.  The
    lines across the plot are the 
    boundary of the classically allowed region $E_0(N)$ and
    the bifurcation line $E_1(N)$.}
\label{fig16}
\end{figure}

The quantum mechanics of two degrees of freedom is a convenient
testing ground for checking the semiclassical
methods~\cite{Bonini:1999cn,Bonini:1999kj} and, in particular, our
regularization technique. The solutions to the full stationary
Schr\"odinger equation may be found numerically in this case at finite
values of the semiclassical parameter~$\lambda$, and the results may
be extrapolated to $\lambda\to 0$. The suppression factor may be then
compared to the semiclassical result. We performed this check in the
region $E > E_S = 1$, which is most interesting for our purposes.  We
applied the numerical techniques for solving the full stationary
Schr\"odinger equation, which were developed in
Ref.~\cite{Bonini:1999kj}; our results agree with
Ref.~\cite{Bonini:1999kj} where comparison is possible.  The results
of the full quantum mechanical calculation of the suppression exponent
$F$ in the limit $\lambda\to 0$ are represented by points in
Fig.~\ref{fig16}.  The lines in that figure represent the values, for
constant $N$, of the semiclassical exponent $F(E,N)$, which we
obtained by making use of the regularization procedure and
extrapolation to $\epsilon \to 0$.  We see that in the entire
forbidden region (including the region $E > E_1(N)$) the semiclassical
result for $F$ coincides with the exact one.

\subsection{\label{sec:4.3}Classically allowed region}

Finally, let us show that our regularization procedure enables one to
obtain a subset of classical over--barrier solutions, that exist at
high enough energies. This subset is interesting, as it extends all
the way to the boundary of the classically allowed region, $E =
E_0(N)$. In principle, finding this boundary is purely a problem of
classical mechanics, and, indeed, in mechanics of two degrees of
freedom one obtains this boundary numerically by solving numerically
the Cauchy problem for given $E$ and $N$ and all possible oscillator
phases, see Sec.~\ref{sec:over-barrier}.  However, if the number of
degrees of freedom is much larger, this classical problem becomes
quite complicated, as one has to span a high-dimensional space of
Cauchy data.  As an example, a stochastic Monte Carlo technique was
developed in Ref.~\cite{Rebbi:1995zw} to deal with this problem in
field theory context.  The approach below may be viewed as an
alternative to the Cauchy methods.

First, let us recall that all classical over--barrier solutions with
given energy and excitation number satisfy the $T/\theta$ boundary
value problem with $T = 0$, $\theta = 0$.  We cannot reach the allowed
region of $E-N$ plane without regularization, because we have to cross
the line $E_0(N)$ corresponding to the excited sphaleron
configurations in the final state.  However, the excited sphalerons no
longer exist among the solutions of the regularized boundary value
problem at any finite value of $\epsilon$.  This suggests that the
regularization enables one to enter the classically allowed region
and, after taking an appropriate limit, obtain classical solutions
with finite values of $E$, $N$.

By definition, the classically allowed transitions have $F = 0$.
Thus, one expects that in the allowed region, the regularized problem
has the property that 
$F_{\epsilon}(E,N) = \epsilon f(E,N) +
O(\epsilon^2)$.  In view of the inverse Legendre formulas 
(\ref{T}),~(\ref{theta}) the values of $T$ and $\theta$ must be of 
order $\epsilon$: $T
= \epsilon\tau(E,N)$, ${\theta = \epsilon
  \vartheta(E,N)}$, where the quantities $\tau$ and
$\vartheta$ are related to the initial energy and excitation number
(see Eq.~(\ref{E}),~(\ref{N})) in the following way,
\begin{eqnarray}
  E &=& \lim\limits_{\epsilon \to 0} \frac{\partial}{\partial T} 2
  \Imag  S_{\epsilon} = 
  \frac{\partial}{\partial \tau} 2 T_\mathrm{int}(\tau, \vartheta)
  \label{Ecl}
  \;,\\
  N &=& \lim\limits_{\epsilon \to 0} \frac{\partial}{\partial \theta} 2
  \Imag  S_{\epsilon} = 
  \frac{\partial}{\partial \vartheta} 2 T_\mathrm{int}(\tau, \vartheta) ,\label{Ncl}
\end{eqnarray}
and we have used the form~\eqref{Uintnew} of the regularized potential.
Therefore, one expects that one
can invade the classically allowed region by taking a fairly sophisticated limit
$\epsilon \to 0$ with  $\tau \equiv T/\epsilon = \mathrm{const}$,
$\vartheta\equiv \theta/\epsilon = \mathrm{const}$.
In the allowed region the parameters $\tau$ and $\vartheta$ are 
analogous to $T$ and $\theta$.  

By solving the regularized $T/\theta$ boundary 
value problem one
constructs a single solution for given $E$ and $N$.  
On the other hand, for $\epsilon = 0$ there are more classical
over--barrier solutions: they form a  continuous family labeled by  
the initial oscillator phase.
Thus, after taking the limit $\epsilon \to 0$ one obtains a subset of
 over--barrier solutions,  which should therefore obey
some additional constraint.  It is almost obvious, that this constraint is 
that the interaction time $T_{\mathrm{int}}$, Eq.~(\ref{Tint}), is minimal.  
This is shown in App.~\ref{app:B}.

The  subset of classical over--barrier solutions obtained in
the limit ${\epsilon \to 0}$ of the regularized $T/\theta$ procedure
extends all the way to the boundary of the classically allowed region.
Let us see what happens when one approaches this boundary from the classically 
allowed side. At the boundary $E_0(N)$, the unregularized solutions tend to 
excited sphalerons, so the interaction time $T_\mathrm{int}$ is infinite.
This is consistent with~(\ref{Ecl}),~(\ref{Ncl})  only if $\tau$ and 
$\vartheta$ become infinite at the boundary. 
Thus, to obtain a point of the boundary one takes the further limit,
\begin{equation*}
  \big( E_0(N),\, N \big) =
  \lim_{\stackrel{\scriptstyle \tau/\vartheta = \mathrm{const}}
                 {\scriptstyle \tau \to +\infty}}
  \big( E(\tau,\vartheta), N(\tau,\vartheta) \big) \;.
\end{equation*}
Different values of $\tau/\vartheta$ correspond to different points of the
line $E_0(N)$.
In this way one finds  the boundary of the classically
allowed region without an initial-state simulation.  

\begin{figure}
\begin{center}
  \includegraphics[width=\columnwidth]{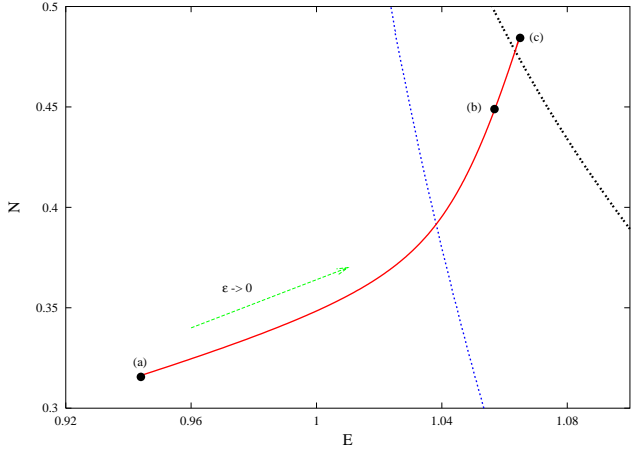}
\end{center}
  \caption{The line $\tau \equiv T/\epsilon = \mathrm{const} = 380$,
    $\vartheta \equiv \theta/\epsilon = \mathrm{const} = 130$
    connecting the points (a): $T = 3.8$, $\theta = 1.3$, $\epsilon =
    0.01$ and (c): $T = 0$, $\theta = 0$, $\epsilon = 0$.  Point (b)
    in the middle has $T=1.82$, $\theta=0.62$, $\epsilon=0.0048$.  As
    the values of $\tau$ and $\vartheta$ are quite large, the limiting
    point is close to the boundary of the classically allowed region.
    The reflection boundary and the boundary of the classically
    allowed region are shown by dashed lines.}
\label{fig17}
\end{figure}

\begin{figure}
  \begin{center}
    \includegraphics[width=\columnwidth]{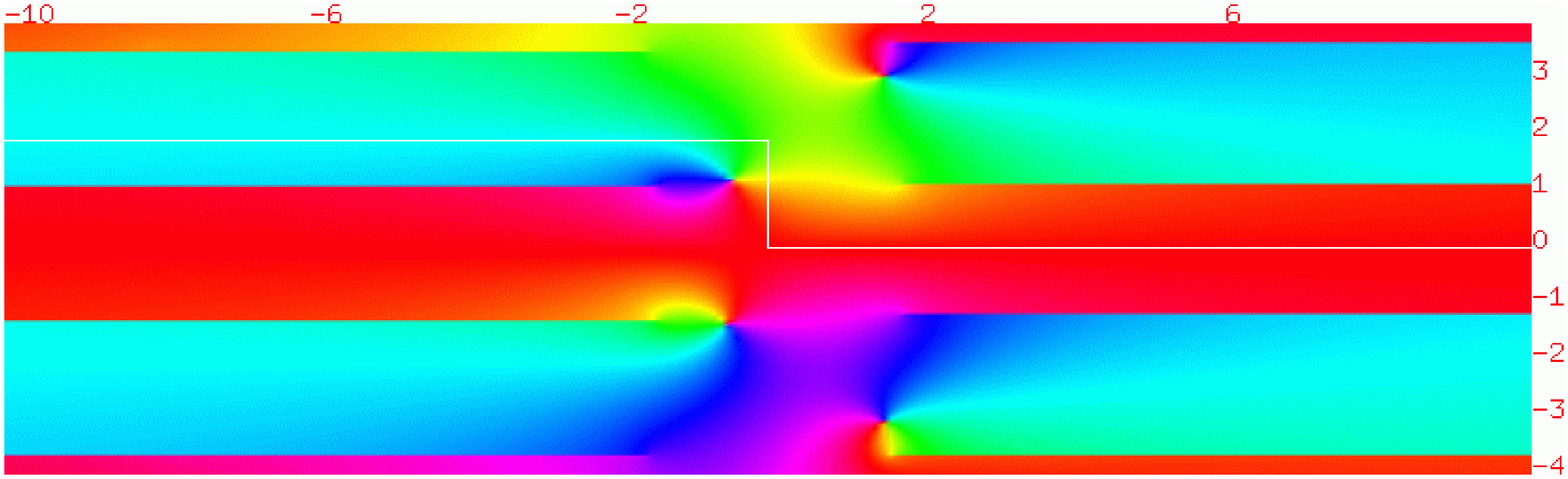}\\
    (a)\\
    \includegraphics[width=\columnwidth]{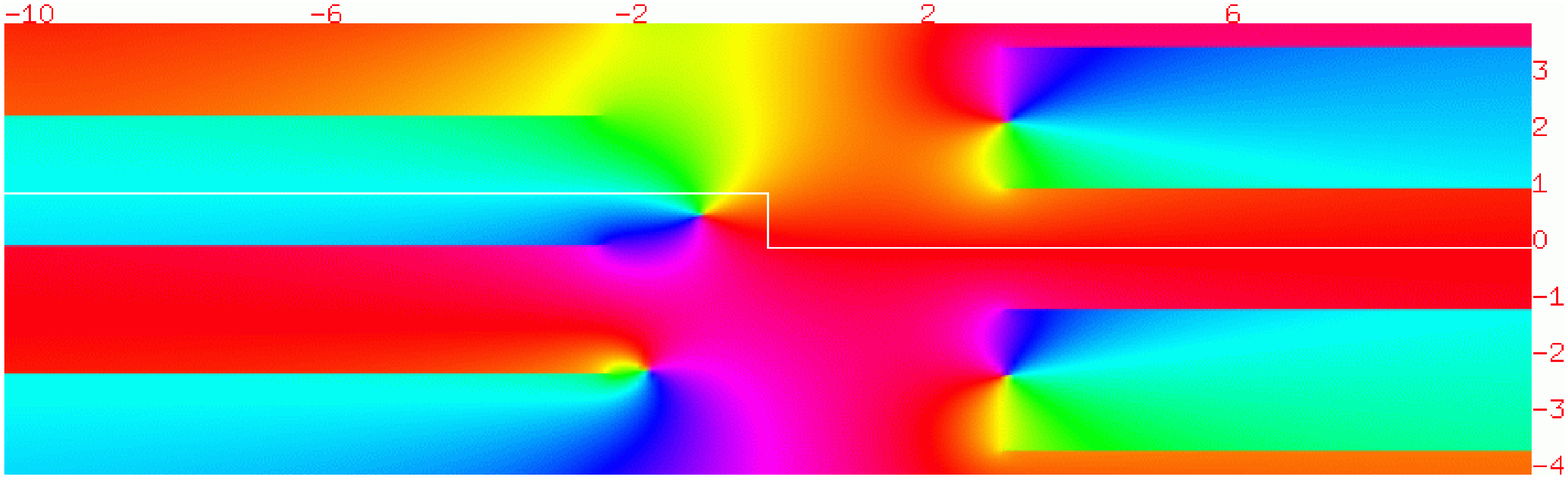}\\
    (b)\\
    \includegraphics[width=\columnwidth]{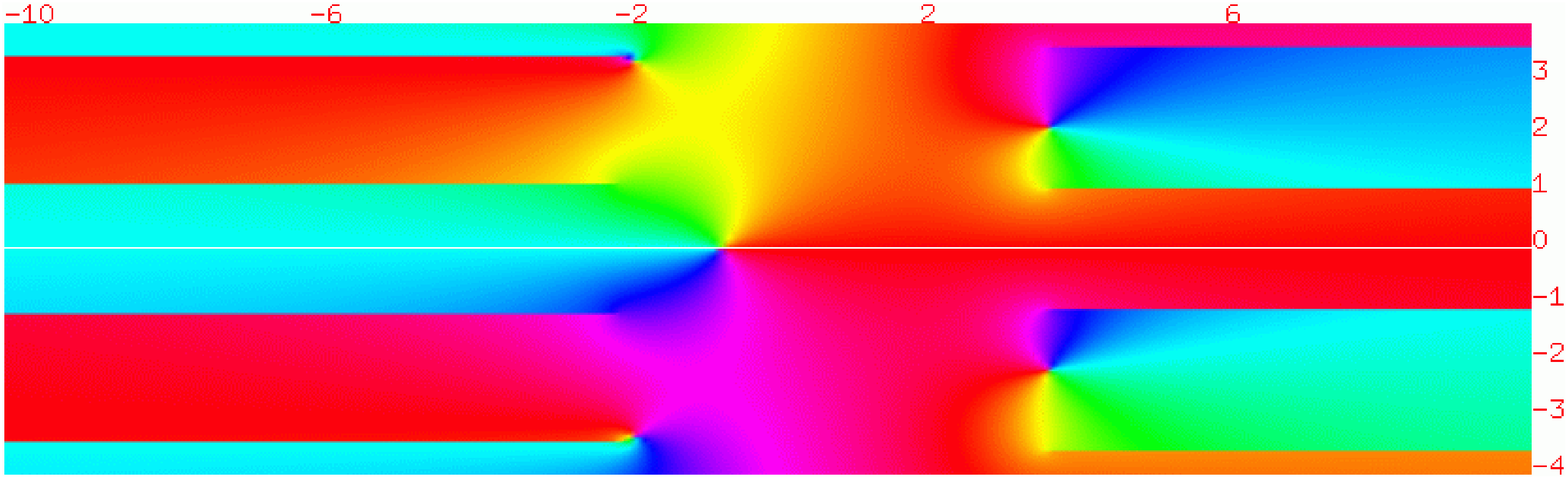}\\
    (c)\\
  \end{center}
  \begin{flushleft}
  $\begin{array}{c}\pi\\-\pi\end{array}
  \begin{array}{c}\includegraphics[width=1cm]{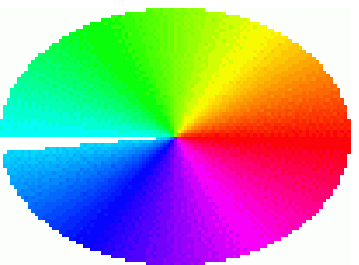}\end{array}
  0$
  \end{flushleft}
  \caption{The phase of the tunneling coordinate in complex time 
    plane at the three points of the curve $\tau = \mathrm{const}$,
    $\vartheta = \mathrm{const}$ (see Fig.~\ref{fig17}). The 
    asymptotics $X\to -\infty$ and $X\to +\infty$ correspond to
    $\mathrm{arg}(X) = \pi$ and $0$ correspondingly.  The contour in
    the time plane is plotted with white line.}
\label{fig18}
\end{figure}

We have checked this procedure numerically.  The line $\tau =
\mathrm{const}, \vartheta = \mathrm{const}$, with varying $\epsilon$,
is shown in Fig.~\ref{fig17}.  The limit $\epsilon \to 0$ exists
indeed---the values of $E$ and $N$ tend to the point (c) of the
classically allowed region.  The phase of the tunneling coordinate
$X(t)$ in complex time plane is shown in Fig.~\ref{fig18} for the
three points (a), (b) and (c) of the curve Fig.~\ref{fig17}.  The
branch points of the solution\footnote{The phase of the tunneling
coordinate turns by $\pi$ around the branch point.  The points where
the phase of the tunneling coordinate turns by $2\pi$ correspond to
the zeroes of $X(t)$.}, the cuts and the contour are clearly seen on
these graphs.

It is worth noting that the left branch points $t_-^*$ move down as
$T$ and $\theta$ approach zero.  Solutions to the right of the point
(b) (along the curve of Fig.~\ref{fig17}) have left branch point in
the lower complex half-plane, $\Imag t^*_- < 0$.  Therefore, the
corresponding contour may be continuously deformed to the real time
axis.  These solutions still satisfy the reality conditions
asymptotically (see Fig.~\ref{fig14}), but show nontrivial complex
behavior at any finite time.


\begin{figure}
  \begin{center}
    \includegraphics[width=\columnwidth]{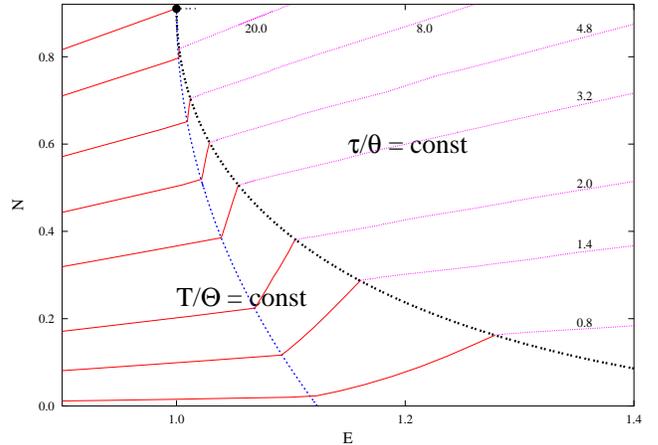}
  \end{center}
  \caption{Lines $T/\theta = \mathrm{const}$ are 
    continuously connected with the lines ${\tau/\vartheta =
      \mathrm{const}}$.  The numbers near the lines give the values of
    $T/\theta = \tau/\vartheta$.}
\label{fig19}
\end{figure}

Making use of the regularized $T/\theta$ procedure, one is able to
approach the boundary of the classically allowed region from both
sides.  The points at this boundary may be obtained by taking the
limits $T \to 0,\; T/\theta = \mathrm{const}$ of the tunneling
solutions and $\tau \to +\infty$, ${\tau/\vartheta = \mathrm{const}}$
of the classically allowed ones.  As $\tau^* \equiv \tau/\vartheta =
T/\theta$ by construction, the lines $\tau^* = \mathrm{const}$ are
continuous at the boundary $E_0(N)$, though may have discontinuity of
the derivatives.  These lines are shown in Fig.~\ref{fig19}.  We see
that the variable $\tau^*$ can be used to parametrize the curve
$E_0(N)$.

For completeness, the lines $\tau = \mathrm{const}$ and $\vartheta =
\mathrm{const}$ in the classically allowed region are presented in
Figs.~\ref{fig12},~\ref{fig13}.

\section{\label{sec:conclusions}Conclusions and Discussion}

Though the semiclassical processes studied in this paper were dubbed
``tunneling'', our results suggest somewhat different interpretation
at energies exceeding the minimum height of the barrier, $E_S$.  At
these energies, it is energetically allowed that the system jumps
above the barrier, and restarts its classical evolution from the
region near its saddle point. This is precisely what we have
observed. From the physical viewpoint, this is not quite what is
normally meant by ``tunneling through a barrier''. Yet the transitions
remain exponentially suppressed (until the energy reaches the boundary
of the allowed region), but, intuitively, the reason is different: to
jump above the barrier, the system has to undergo considerable
rearrangement, unless the incoming state is chosen properly
(i.e.\ unless the oscillator excitation number in our model is large enough).
This rearrangement costs exponentially small probability factor. We
note that similar exponential factor was argued to appear in various
field theory processes with multi-particle final
states~\cite{Banks:1990zb,Zakharov:1991rp,Veneziano:1992rp,Rubakov:1995hq}.

The $T/\theta$ boundary value problem, equipped with our regularization 
technique, enables one to deal with this situation,
which, in the language of complex classical solutions,
corresponds to a new physically relevant branch.
The regularization procedure chooses automatically the correct 
branch, and thus provides an efficient way to cross the bifurcation lines
where different branches merge. The phenomenon discussed in this paper 
appears to be quite general, as it exists in diverse quantum systems, from 
two-dimensional quantum mechanics to quantum field theory. The appropriate 
generalization of our regularized $T/\theta$ procedure is straightforward;
we are going to report on this application to baryon--number violating 
processes in gauge theory in future publication~\cite{BLRRT:prepare}.

\begin{acknowledgments}
The authors are indebted to V.~Rubakov and C.~Rebbi for numerous
valuable discussions and criticism, A.~Kuznetsov and S.~Sibiryakov for
helpful discussions, and S.~Dubovsky, D.~Gorbunov, A.~Penin and
P.~Tinyakov for stimulating interest.  We wish to thank Boston
University's Center for Computational Science and Office of
Information Technology for allocations of supercomputer time.  This
reseach was supported by Russian Foundation for Basic Research grant
02-02-17398, U.S.  Civilian Research and Development Foundation for
Independent States of FSU~(CRDF) award RP1-2364-MO-02, and under DOE
grant US DE-FG02-91ER40676.
\end{acknowledgments}

\appendix

\section{\label{app:A}$T/\theta$ boundary value problem}

The semiclassical method for calculating the probability of
tunneling from a state with a few parameters fixed was developed in
\cite{Rubakov:1992ec, Rubakov:1992fb, Kuznetsov:1997az,
Bezrukov:2001dg} in context of field theoretical models and in
\cite{Bonini:1999kj, Bonini:1999cn} in quantum mechanics.  Here
we outline the method, adapted to our model of two degrees of freedom.

\subsection{\label{app:A.1}Path integral representation of the
transition probability}

We begin with the path integral representation for the probability 
of tunneling  from the asymptotic region 
$X\to -\infty$ through a potential barrier.
Let the incoming state $|E,\;N\rangle_\delta$ have fixed energy and oscillator 
excitation number, and has support only for $X\ll 0$, well outside the range 
of the potential barrier. 
The inclusive tunneling probability for states of this type is given by
\begin{multline}
\mathcal{T}(E,N) 
= \lim\limits_{t_f - t_i \to \infty} \Bigg\{ \int\limits_0^{+\infty}
dX_f \int\limits_{-\infty}^{+\infty} dy_f \\
\left|\langle X_f,y_f|\mathrm{e}^{-i\hat{H}(t_f - t_i)}|E,N\rangle
\right|^2\Bigg\}\;.
\end{multline}
This probability can be reexpressed in terms of the
transition amplitudes 
\begin{equation}
{\cal A}_{fi} = \langle X_f,\;y_f| \mathrm{e}^{-i H(t_f - t_i)} |X_i,\;y_i\rangle
\end{equation}
and initial-state matrix elements
\begin{equation}
{\cal B}_{ii'} = \langle X_i,\;y_i| E,\;N\rangle
\langle E,\;N| X_i',\;y_i'\rangle
\end{equation}
in the following way,
\begin{multline}\label{Ttemp}
{\cal T}(E,\;N) = \lim\limits_{t_f-t_i \to \infty}\Bigg\{
\int\limits_{0}^{+\infty} dX_f \int\limits_{-\infty}^0
d{X}_i\,d{X}_i'\\
\int\limits_{-\infty}^{+\infty} dy_i\,dy_i'\,dy_f\;
 {\cal A}_{fi} 
{\cal A}_{i'f}^*
{\cal B}_{ii'}\Bigg\}\;.
\end{multline}
The transition amplitude and its complex conjugate have the familiar path integral 
representation:
\begin{eqnarray}\label{A}
  \mathcal{A}_{fi} &=& \int [d\vec{x}]
  \Bigg|_{\stackrel{
    \scriptscriptstyle\vec{x}(t_i) = \vec x_i}{
    \scriptscriptstyle\vec{x}(t_f) = \vec x_f}}
  \e^{iS[\vec{x}]}
  \;,\\
  \mathcal{A}_{i'f}^* &=& \int [d\vec{x'}]\Bigg|_{\stackrel{
    \scriptscriptstyle\vec{x}'(t_i) = \vec x'_i}{
    \scriptscriptstyle\vec{x}'(t_f) = \vec x_f}}
  \e^{-iS[\vec{x'}]}
  \;,\nonumber
\end{eqnarray}
where $\vec{x} = (X,y)$, and  $S$ is the action of the model.  To obtain a similar  
representation for the initial-state matrix elements let 
us rewrite ${\cal B}_{ii'}$ as follows,
\begin{equation}
{\cal B}_{ii'}  = 
\langle X_i,\;y_i| \hat{P}_E \hat{P}_N|X_i',\;y_i'\rangle\;,
\end{equation}
where $\hat{P}_N$ and $\hat{P}_E$ denote the projectors onto 
states with oscillator excitation number $N$ and total energy 
$E$ respectively.  
It is convenient to use the coherent state formalism for the 
$y$-oscillator and choose the momentum basis for the $X$-coordinate.
In this representation, the 
kernel of the projector operator $\hat{P}_E\hat{P}_N$ takes the form
\begin{multline*}
\langle q,\;b| \hat{P}_E \hat{P}_N | p,\;a\rangle = \frac{1}{(2 \pi)^2}
\int d\xi\,d \eta \\
\mathrm{exp}\left(- iE\xi  - iN\eta + \frac{i}2 p^2\xi + 
\mathrm{e}^{i \omega \xi + i\eta} \bar{b} a\right)\delta(q - p)\;,
\end{multline*}
where $|p,\;a\rangle$ is the eigenstate of the 
center-of-mass momentum $\hat{p}_X$ and $y$-oscillator annihilation 
operator $\hat{a}$ with eigenvalues $p$ and $a$ 
respectively.
It is straightforward to express this matrix element in the 
coordinate representation using the formulas
\begin{eqnarray*}
\langle y | a \rangle  &=&  \sqrt[4]{\frac\omega\pi} 
\mathrm{e}^{ - \frac12 a^2 + 
\sqrt{2 \omega} a y - \frac12 \omega y^2}\;,\\
\langle X | p \rangle  &=&  \frac{1}{\sqrt{2 \pi}} \mathrm{e}^{i p X}\;.
\end{eqnarray*}
Evaluating  the Gaussian integrals over $a$, $b$, $p$, $q$, we obtain
\begin{eqnarray}
\label{B}
{\cal B}_{ii'} &=& \int d\xi \; d\eta\;  \mathrm{exp}\left(
- i E \xi - i N \eta - \frac{i}{2} \frac{(X_i - X_i')^2}{\xi} \right.\nonumber\\
&&+\left.\frac{\omega}{1 - \mathrm{e}^{- 2 i \omega\xi - 2 i \eta}}\left[
\frac{y_i^2 + y_i^{\prime 2}}{2}(1 + \mathrm{e}^{- 2 i \omega \xi - 2 i \eta})
 \right.\right.\nonumber\\
&&\quad\quad\quad\quad\quad\quad\quad\quad\quad -2 y_i y_i' \mathrm{e}^{- i \omega \xi - i \eta}\bigg]\bigg)
\end{eqnarray}
where we omit the pre-exponential factor depending on $\eta,\;\xi$.
For the subsequent formulation of the boundary value problem it is convenient 
to introduce the notation
\begin{equation*}
  T = - i \xi\;,\qquad  \theta = - i \eta\;.
\end{equation*}
Then, combining together the integral representations (\ref{B}) and~(\ref{A})
and rescaling coordinates, energy and excitation number $\vec{x} \to 
\vec{x}/\sqrt{\lambda}$, $E\to E/\lambda$, $N\to N/\lambda$, we finally
obtain
\begin{multline}
\label{TPI}
{\cal T}(E,\;N) = \lim\limits_{t_f-t_i \to \infty} \Bigg\{
\int\limits_{-i\infty}^{+i\infty}  dT\;d\theta \int \;
[d\vec{x}\;d\vec{x}'] \\
\exp\left\{- \frac{1}{\lambda}F[\vec{x},\;\vec{x}';\;T,\;\theta]\right\}\Bigg\}\;,
\end{multline}
where 
\begin{multline}
F[\vec{x},\;\vec{x}';\;T,\;\theta] = - i S[X,y] + i S[X',\;y'] \\
- E T - N \theta + B_i(\vec{x}_i, \vec{x'}_i;
T,\theta).\label{FF}
\end{multline}
Here the non-trivial initial term $B_i$ is 
\begin{align}\label{Bi}
  B_i = \ & \frac{(X_i - X_i')^2}{2 T}
  \nonumber \\
    & - \frac{\omega}{1 - \mathrm{e}^{2 \omega T + 2 \theta}}\bigg[
      \frac12(y_i^2 + {y_i'}^2)(1 + \mathrm{e}^{2 \omega T +  2 \theta})
  \nonumber\\
    & \hphantom{- \frac{\omega}{1 - \mathrm{e}^{2 \omega T + 2 \theta}}\bigg[}
      - 2 y_i y_i' \e^{\omega T + \theta}\bigg]\;.
\end{align}
In~\eqref{TPI} $\vec{x}$ and $\vec{x'}$ are independent integration
variables, while $\vec{x}_f'\equiv \vec{x}_f$, see Eqs.~\eqref{Ttemp}.

\subsection{\label{app:A.2}The boundary value problem}

For  small $\lambda$, the path integral \eqref{TPI}
is dominated by a stationary point of the functional $F$. 
Thus, to calculate the tunneling exponent, we extremize this 
functional with respect to all variables of integration: $X(t)$, $y(t)$, 
$X'(t)$, $y'(t)$, $T$, $\theta$. Note that because of the limit $t_f - t_i 
\to + \infty$, the  variation 
with respect to the initial and final values of coordinates leads to 
boundary conditions imposed at \emph{asymptotic} $t\to \pm\infty$, rather 
than at finite times $t_i,t_f$.
Note also that the  stationary points may be complex.

The variation of the functional \eqref{FF} with respect to the coordinates 
at intermediate times gives second order equations 
of motion, in general complexified,
\begin{subequations}
\label{problem1}
\begin{equation}
\frac{\delta S}{\delta X(t)} = 
\frac{\delta S}{\delta y(t)} = 
\frac{\delta S'}{\delta X'(t)} = 
\frac{\delta S'}{\delta y'(t)} = 0 \;. \label{equation}
\end{equation}
The boundary conditions at the final time $t_f \to +\infty$ are obtained by 
extremization 
of $F$ with respect to $X_f\equiv X_f'$, $y_f\equiv y_f'$.
These are 
\begin{equation}\label{bcf}
  \dot{X_f} = \dot{X_f'}
  \;,\qquad
  \dot{y_f} = \dot{y_f'}
  \;.
\end{equation}
It is convenient to write the conditions at the initial time (obtained 
by varying $X_i$, $y_i$, $X_i'$, $y_i'$) in terms of the asymptotic quantities.
At the initial moment of time $t_i \to -\infty$, the system moves in the region $X\to -\infty$,
well outside of the range of the potential barrier.  
Equations \eqref{equation} in this region describe free motion of decoupled oscillator,
and the general solution takes the following form,
\begin{eqnarray}
&&X(t) = X_i + p_i(t - t_i)\nonumber,\\
&&y(t) = \frac{1}{\sqrt{2 \omega}} \left[a \mathrm{e}^{ - i \omega(t - t_i)}+
\bar{a} \mathrm{e}^{i \omega(t - t_i)}\right]\nonumber\;,
\end{eqnarray}
while the solution for $X'(t)$, $y'(t)$ has similar form.
For the moment, $a$ and $\bar{a}$ are independent variables.
The initial boundary conditions in terms of the asymptotic variables 
$X_i$, $p_i$, $a$, $\bar{a}$ take the form:
\begin{eqnarray}
&&p_i = p_i' = - \frac{X_i - X_i'}{i T},\nonumber\\
&&a'+\bar{a}' = a\mathrm{e}^{\omega T + \theta} + \bar{a} \mathrm{e}^{- \omega T
- \theta} \label{bci}\;,\\
&&a+\bar{a} = a'\mathrm{e}^{-\omega T - \theta} + \bar{a}' \mathrm{e}^{\omega T
+ \theta} \nonumber\;.
\end{eqnarray}
The variation with respect to the Lagrange multipliers $T$ and $\theta$ 
gives the relation between the values of $E$, $N$ and initial asymptotic 
variables (here we use the boundary conditions \eqref{bci}),
\begin{eqnarray}
  E &=& \frac{p_i^2}{2} + \omega N\label{EN0}\;,\\
  N &=& a\bar{a}\nonumber\;.
\end{eqnarray}
Equations \eqref{equation} -- \eqref{EN0} constitute the complete set of 
saddle-point equations for the functional $F$. 

The variables $X'$ and $y'$ originate from the conjugate
amplitude ${\cal A}_{i'f}^*$ (see Eq.~\eqref{A}), which suggests that
they are the complex conjugate to $X$, $y$. Indeed, the Ansatz ${X'(t) =
X^*(t)}$, ${y'(t) = y^*(t)}$ is compatible with the boundary value
problem \eqref{problem1}. Then the Lagrange multipliers $T$,
$\theta$ are real, and the problem \eqref{problem1} may
be conveniently formulated at the contour ABCD in the complex time
plane (see Fig.~\ref{fig2}).
\end{subequations} 

Now we have only two independent complex variables $X(t)$ and $y(t)$,  
that have to satisfy the classical equations of motion in the interior of the 
contour,
\begin{subequations}
\label{problem2}
\begin{equation}
\frac{\delta S}{\delta X(t)} = 
\frac{\delta S}{\delta y(t)} =0\;. \label{equation2}
\end{equation}
The final boundary conditions (see Eq.~\eqref{bcf}) become the
conditions of the reality of the variables $X(t)$ and $y(t)$ at the
asymptotic part D of the contour:
\begin{eqnarray}
\begin{array}{l}
\Imag X_f = 0,\;\; \Imag y_f = 0\;,\\[1ex] 
\Imag \dot{X}_f = 0\;\;\Imag\dot{y}_f = 0\;,
\end{array} \qquad t\to +\infty\label{bcf2}\;.
\end{eqnarray}
The seemingly  complicated initial conditions \eqref{bci} simplify when written 
in terms of the  time coordinate $t' = t + i T/2$ running along the part AB of 
the contour. Let us again write the asymptotics of a solution, but now 
along the initial part AB of the contour:
\begin{eqnarray}
&&X = X_0 + p_0 (t' - t_i)\;,\nonumber \\ 
&&y = \frac{1}{\sqrt{2 \omega}}\left[u \mathrm{e}^{ - i \omega(t' - t_i)} +
v\mathrm{e}^{i \omega(t' - t_i)}\right]\;.\nonumber
\end{eqnarray}
In terms of $X_0$, $y_0$, $u$ and $v$, the boundary conditions \eqref{bci} are
\begin{eqnarray}
&&\Imag X_0 = 0,\;\; \Imag p_0 = 0\;,\label{bci2}\\
&&v = u^*\mathrm{e}^\theta\;.\nonumber
\end{eqnarray}
\end{subequations} 
Finally, we write Eqs.~\eqref{EN0} in terms 
of the asymptotic variables along the initial part of the contour:
\begin{eqnarray}
&&E = \frac{p_0^2}{2} + \omega N\;,\label{EN}\\
&&N = \omega u v\;.\nonumber
\end{eqnarray}
These equations determine 
the Lagrange multipliers $T,\;\theta$ in terms of $E$, $N$. Alternatively,
we can solve the problem \eqref{problem2} for given values of $T$, $\theta$
and find the values of $E$,$N$ from Eqs.~\eqref{EN}, what is 
more convenient computationally.

Given a solution to the problem \eqref{problem2}, the exponent $F$ 
is the value of the functional~\eqref{FF} at this saddle point. Explicitly
\begin{equation}
F = - ET - N\theta + 2 \Imag S_0(T,\theta)\;, \label{FFF}
\end{equation}
where $S_0$ is the action of the system, integrated by parts
\begin{equation*}
S_0 = \int \!dt \left[-\frac12 X\frac{d^2}{dt^2}X - \frac12 y \frac{d^2}{dt^2}
y - \frac{\omega^2}{2} y^2 - U_\mathrm{int}(X,y)\right].
\end{equation*}
Note that we did not make use of the constraints~\eqref{EN} to obtain
the formula \eqref{FFF}, so we still can extremize \eqref{FFF} with
respect to $T$ and $\theta$ (see discussion in Sec.~\ref{sec:Ttheta}).

The classical problem \eqref{problem2} is conveniently dubbed
$T/\theta$ boundary value problem.  Eqs.~\eqref{bcf2} and \eqref{bci2}
imply eight real boundary conditions for two complex second-order
differential equations \eqref{equation2}.  However, one of these real
conditions is redundant: Eq.~\eqref{bcf2} implies that the (conserved)
energy is real, so the condition $\Imag p_0\to0$ is automatically
satisfied (note that the oscillator energy $E_\mathrm{osc}=\omega
uv=\omega\e^\theta uu^*$ is real).  On the other hand, the system
\eqref{problem2} is invariant under time translations along the real
axis.  This invariance may be fixed, e.g., by demanding that
$\Real(X)$ takes a prescribed value at a prescribed large negative
time $t'_0$ (note that other ways may be used instead.  In particular,
for $E<E_1(N)$ it is convenient to impose the constraint
$\Real\dot{X}({t=0})=0$).  Together with the latter requirement, we
have exactly eight real boundary conditions for the system of two
complexified (i.e.\ four real) second-order equations.

\section{\label{app:B}A property of solutions to $T/\theta$ problem in
the allowed region}

For given $E$, $N$ there is only one over--barrier classical solution
which is  
obtained in the limit $\epsilon \to 0$ of the regularized $T/\theta$ 
procedure. To see what singles out this solution,  let us analyze the regularized functional
\begin{equation}
F_{\epsilon}[q] = F[q] + 2 \epsilon T_{\mathrm{int}}[q],
\end{equation}
where $q$ denotes the variables $\vec{x}(t),\; \vec{x}'(t)$ and 
$T,\;\theta$ together.  
The unregularized functional $F$ has a
valley of extrema $q^{e}(\varphi)$ 
corresponding to different values of the initial oscillator phase $\varphi$.
Clearly, at small $\epsilon$ the extremum of $F_\epsilon$ is close to a point 
in this valley with the phase extremizing $T_\mathrm{int}[q^e(\varphi)]$.
\begin{equation}
  \frac{d}{d\varphi} T_{\mathrm{int}}[q^e(\varphi)] = 0\;.
\end{equation}
Hence, the solution  $q^e_{\epsilon}$ 
of the regularized $T/\theta$ boundary value problem 
tends to the over--barrier classical solution, 
with $T_\mathrm{int}$ extremized with respect to 
the initial oscillator phase.  

Because $U_{\mathrm{int}}(\vec{x}) > 0$, $T_{\mathrm{int}}$ is a positive 
quantity with
at least one minimum.  In normal 
situation there is only one saddle point of $F_\epsilon$, so by solving 
the  $T/\theta$ boundary value problem one obtains the classical solution
with the time of interaction minimized.

\begin{figure}
  \begin{center}
    \bigskip
    \includegraphics[width=0.8\columnwidth]{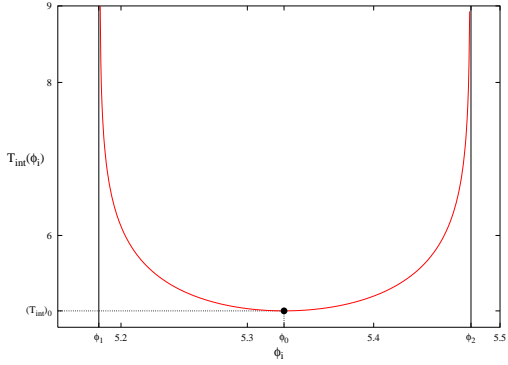}
    \vspace{-4ex}
  \end{center}
  \caption{The dependence of the interaction time 
    $T_{\mathrm{int}}$ on the initial oscillator phase at fixed values of 
    $E$ and $N$, corresponding to the point (c) in Fig.~\ref{fig18}.
    The black point marks the initial oscillator phase and interaction time
    calculated on the solution (c).}
\label{fig20}
\end{figure}

The dependence of the interaction time on the initial oscillator
phase for given values of energy and excitation number is shown
in Fig.~\ref{fig20}.  We see that $T_\mathrm{int}$ indeed has one minimum,
and the corresponding solution indeed coincides with the limit of the solution 
of the regularized 
$T/\theta$ boundary value problem (black point on the graph).

\bibliographystyle{h-physrev4}
\bibliography{slac,paper}

\begin{thebibliography}{10}

\bibitem{Kuznetsov:1997az}
A.~N. Kuznetsov and P.~G. Tinyakov,
\newblock Phys. Rev. {\bf D56}, 1156 (1997), [hep-ph/9703256].

\bibitem{Bezrukov:2001dg}
F.~Bezrukov, C.~Rebbi, V.~Rubakov and P.~Tinyakov,
\newblock hep-ph/0110109.

\bibitem{Miller}
W.~H. Miller,
\newblock Adv. Chem. Phys. {\bf 25}, 69 (1974).

\bibitem{'tHooft:1976fv}
G.~'t~Hooft,
\newblock Phys. Rev. {\bf D14}, 3432 (1976).

\bibitem{Ringwald:1990ee}
A.~Ringwald,
\newblock Nucl. Phys. {\bf B330}, 1 (1990).

\bibitem{Espinosa:1990qn}
O.~Espinosa,
\newblock Nucl. Phys. {\bf B343}, 310 (1990).

\bibitem{BLRRT:prepare}
F.~L. Bezrukov, D.~Levkov, C.~Rebbi, V.~A. Rubakov and P.~Tinyakov,
\newblock In preparation  (2003).

\bibitem{Bonini:1999kj}
G.~F. Bonini, A.~G. Cohen, C.~Rebbi and V.~A. Rubakov,
\newblock Phys. Rev. {\bf D60}, 076004 (1999), [hep-ph/9901226].

\bibitem{Bonini:1999cn}
G.~F. Bonini, A.~G. Cohen, C.~Rebbi and V.~A. Rubakov,
\newblock quant-ph/9901062.

\bibitem{Klinkhamer:1984di}
F.~R. Klinkhamer and N.~S. Manton,
\newblock Phys. Rev. {\bf D30}, 2212 (1984).

\bibitem{Rubakov:1992ec}
V.~A. Rubakov, D.~T. Son and P.~G. Tinyakov,
\newblock Phys. Lett. {\bf B287}, 342 (1992).

\bibitem{Rubakov:1992fb}
V.~A. Rubakov and P.~G. Tinyakov,
\newblock Phys. Lett. {\bf B279}, 165 (1992).

\bibitem{Bonini:1999fc}
G.~F. Bonini {\em et~al.},
\newblock hep-ph/9905243.

\bibitem{Manton:1983nd}
N.~S. Manton,
\newblock Phys. Rev. {\bf D28}, 2019 (1983).

\bibitem{Khlebnikov:1991th}
S.~Y. Khlebnikov, V.~A. Rubakov and P.~G. Tinyakov,
\newblock Nucl. Phys. {\bf B367}, 334 (1991).

\bibitem{Rebbi:1995zw}
C.~Rebbi and J.~Robert~Singleton,
\newblock hep-ph/9502370.

\bibitem{Banks:1990zb}
T.~Banks, G.~Farrar, M.~Dine, D.~Karabali and B.~Sakita,
\newblock Nucl. Phys. {\bf B347}, 581 (1990).

\bibitem{Zakharov:1991rp}
V.~I. Zakharov,
\newblock Phys. Rev. Lett. {\bf 67}, 3650 (1991).

\bibitem{Veneziano:1992rp}
G.~Veneziano,
\newblock Mod. Phys. Lett. {\bf A7}, 1661 (1992).

\bibitem{Rubakov:1995hq}
V.~A. Rubakov,
\newblock hep-ph/9511236.

\end{thebibliography}

\end{document}